\documentstyle[preprint,eqsecnum,aps]{revtex}

\newcommand{\fav}{\langle f \rangle}  
\newcommand{\favn}{\langle f(n) \rangle}  
  
\newcommand{\favav}{\langle\langle f \rangle\rangle}

\begin{document}
\draft
%\preprint{STH-98-01}
\title{Phenomenology \\
of Globally Coupled Map Lattice and its Extension}
\author{Tokuzo Shimada}
\address{Department of Physics, School of Science and Technology, Meiji University\\
Higashi-Mita 1-1, Kawasaki, Kanagawa 214, Japan
}
\date{\today}
\maketitle
\begin{abstract}
We revisit the globally coupled map lattice (GCML) and also propose a new extended globally coupled map 
lattice (EGCML) with an inverse power law interaction.
In GCML we clarify the mechanism of the basic posi-nega switch in the two-cluster regime.
We show that there is a natural mechanism in GCML which guarantees no mixing of maps
across their mean field in the chaotic transient process. In the turbulent regime of GCML 
there is a prominent period three window. In the correlator analysis we also find a remnant 
of periodic motion of quasi-clusters with high rate mixing almost everywhere in the turbulent
regime of the large-size GCML.  The so-called hidden coherence is the most modest remnant.
The EGCML shares the same intriguing properties with GCML 
and exhibits an amazing spatial cluster formation during the chaotic transient process.   
An analytic approach is proposed which relates the periodicity manifestation in the 
turbulent regime of GCML to the periodic window of a single logistic map.\\

\end{abstract}
%%%%%
\section{Introduction}

Recently there has been much progress in the study of synchronization of 
nonlinear maps \cite{ka,kb,kc,sinha} and flows \cite{pecora,kowalski,noise,fns,fs,pikovskya}.
This may lead to the clarification of the intelligence supposed to come from 
the synchronization among the neurons in the neural network, e.g.\cite{yorke}.
Especially the globally coupled map lattice (GCML), a system of identical maps 
coupled via their mean field, has surprisingly rich clustering phases in the space of 
the nonlinearity parameter and the coupling. 
In the ordered two-cluster phase, GCML exhibits a posi-nega switch between clusters 
of synchronizing maps controlled by external inputs\cite{ka}.
Furthermore even in the turbulent regime there emerges some hidden correlation 
which was first observed as the violation of the law of large numbers (LLN) in the 
fluctuation of the mean field in the time series\cite{kb,pikovskyb}.
GCML is a natural extension of the spin-glass models\cite{biology,spinglass} and these interesting 
features may be relevant to the intelligence activity. 
All the maps in GCML contribute to the mean field with an equal weight and 
hence there is no notion of distance between maps. It is conjectured that 
the intriguing properties of GCLM, both the posi-nega switch and the hidden 
coherence, come from this bona fide scaling invariance of GCML\cite{ka,kb}. 
However, neither the reason why there occurs complete swapping of the maps between 
clusters in the posi-nega switch in the two-cluster phase nor the global feature 
of the turbulent phase has been elucidated so far.
The purpose of this article is twofold. 
Firstly, we revisit GCML. We show that there is a natural mechanism which guarantees 
the complete swapping of maps in the two-cluster posi-nega switch. 
The key to this mechanism is the motion of the mean field during the transient process. 
We show that even though whole maps are released from tightly synchronizing clusters to 
chaotic violent movements during the transient process, still they never mix across their mean field. 
We clarify the mechanism for this phenomenon. 
We also perform an extensive survey of the turbulent regime of GCML. 
We show that this regime has a prominent period three window and is full of periodicity remnants.
A correlator analysis detects quasi-periodic clusters with high mixing rate among them almost everywhere
in the turbulent regime of the large-size GCML. The so-called hidden coherence occurs in-between the
regions of the quasi-clusters and the valleys with no violation of LLN as the most modest appearance 
of the periodicity remnants.

Secondly, we construct a new extended globally coupled map lattice (EGCML). Our EGCML is a 
two-dimensional map lattice and the maps are under all-to-all interaction as in the GCML. 
But instead of the uniform interaction in GCML, the interaction in EGCML obeys an inverse power law with
respect to the distance between the maps on the lattice.
EGCML locates at an intermediate position between the nearest neighbor coupling map model 
and the GCML with uniform interaction. 
Thus EGCML facilitates a test whether the salient feature of GCML is due to the scale invariant interaction in it.
We find that EGCML indeed inherits the intriguing properties of GCML though 
they turn out diffuse to some extent. For instance, in the two-cluster phase of EGCML, the two clusters 
follow the same bifurcation tree with that of GCML 
when the population unbalance between the clusters are increased but the synchronization of maps in each 
cluster is looser in EGCML than in GCML.  We also observe in the turbulent regime of the EGCML the same
periodic window as well as the quasi-periodic clusters with high mixing rate. 

GCML exhibits the basic posi-nega switch as well as more intricate controlled switches 
among coded clusters at various hierarchical levels in its simplest construction\cite{ka} and 
certainly serves as a model for intelligence activity. 
But the very fact that it is a system of globally coupled maps with an equal coupling makes it lose 
the notion of the distance between maps. In a way it is a zero dimensional chaotic field theory.
Apart from the intelligence one is interested to know what is the (classical) field theory of chaotic maps 
on the lattice at criticality. In this aspect the synchronizing maps on the lattice may be regarded as 
the analog of Higgs condensation in the field theory\cite{fns,fs}. 
We can use EGCML to see the general trend how the spatial clustering of maps occurs under the scale 
invariant interaction on the lattice. We find that EGCML does exhibit a self-organized formation of 
the spatial clusters (condensation) due to the synchronization between the maps during the chaotic 
transient process in the switch and after it.  
   In the EGCML switch the positive (negative) cluster becomes negative (positive) evolving in the same 
   two-cluster attractor as before with only a few percent change of the population unbalance between the 
   two clusters in most of the runs. However,  there occurs the mixing of maps between the clusters. 

The organization of this article is as follows. In Sec. II we pay attention to two important rules 
in the GCML. These are the reciprocity of the couplings among the maps and an invariance rule that
the maps do not change their mean field under the interaction.  We then construct EGCML based on these rules.
In Sec. III we compare the phase diagrams of both models and show that they  are almost the same each other.
In Sec. IV we analyze the two-cluster regime of both models. 
For the GCML we clarify how the basic posi-nega switch between two clusters in the two-cluster regime 
can be realized. We then turn to EGCML and investigate the switch process in it. 
We show that amazing spatial clusters are formed during the chaotic transient process in the switch.
In Sec. V we present the results of an extensive statistical analysis over the whole turbulent regime
of both GCML and EGCML.  We show that even in the turbulent regime the maps are under coherence
with various manifestations of periodicity effect. 
We conclude in Sec. VI.

%%%%%
\section{Model Construction}
The coupled map lattice model in general is a system of one-dimensional field
on a lattice $\Lambda$ with number of sites $N$ which evolves simultaneously 
by an iteration of the following two steps:  
\begin{eqnarray}
	x_P (n) &\longmapsto& f(x_P (n)) \nonumber \\
	f(x_P (n)) &\longmapsto& x^\prime_P\equiv x_P(n+1) 
	= (1-\varepsilon)f(x_P)+\varepsilon\sum_{Q\in\Lambda}J_{PQ}f(x_Q).
	~~~~ \forall P \in \Lambda. \label{generic}
\end{eqnarray}
In the first step all $x_P$ on the lattice (called as \lq map\rq~ conventionally) are simultaneously 
mapped by a nonlinear function $f$.  
The function $f$ could be different site by site but we consider here the simplest case that 
$f$ is a simple logistic map $f(x)=1-a x^2$ common to all sites.
In the second step the maps undergo interaction between themselves. 
The $N\times N$ constant matrix $(J_{PQ})$ represents 
the coupling between the maps at ${P}$ and ${Q}$ on the lattice and 
the parameter $\varepsilon$ is an overall coupling constant.
Thus the model in the simplest form has only two parameters, the nonlinear 
parameter $a$ and the coupling constant $\varepsilon$. 
If we rewrite (\ref{generic}) in terms of $y_P\equiv f(x_P)$,  we obtain 
\begin{eqnarray}
	y^\prime_P=f\left( (1-\varepsilon)y_P
	+\varepsilon\sum_{Q\in\Lambda}J_{PQ}y_Q\right) \label{network},
\end{eqnarray}
which is a familiar form in the neural network analysis \cite{biology,spinglass}.

\subsection{The globally coupled map lattice: GCML}
In GCML \cite{ka} we consider N maps $x_P (P \in \Lambda)$ which evolve as  
\begin{eqnarray}
	x^\prime_P=(1-\varepsilon)f(x_P)+\varepsilon \fav~, \label{global1}
\end{eqnarray}
where  $\fav$ is the mean field on the lattice  
\begin{eqnarray}
	  \fav \equiv \frac{1}{N} \sum_{Q\in\Lambda}f(x_Q). \label{global2}
\end{eqnarray}
Thus the coupling matrix of GCML is simply
\begin{eqnarray}
	J_{PQ}=\frac{1}{N}, ~~~~~  P,Q \in \Lambda 
\end{eqnarray}
and clearly satisfies the reciprocity
\begin{eqnarray}
	\mbox{(I)} ~~~~~J_{PQ}=J_{QP} \label{reciprocity}.
\end{eqnarray}
From (\ref{global1}) and (\ref{global2}) we can also derive a relation  
\begin{eqnarray}
	\mbox{(II)}~~~~\sum_{P\in \Lambda} {x^\prime_{P}}=
	\sum_{P\in \Lambda} f(x_{P}). \label{sumrule}
\end{eqnarray}
This relation guarantees that the mean field value is kept invariant when the maps undergo the interaction
at the second step of the interaction.
These two relations serve as a guide to construct the map lattice models in general.
Let us examine the implication of the invariance rule (II).
The nonlinearity of $f$ generally magnifies the variance among the maps. 
Therefore, the first step may be regarded as a {\it defocusing lens} with 
some {\it random prisms} installed in it.
The larger the parameter $a$ is, the more strongly the defocusing lens acts on the system of maps. 
In the second step every $f(x_P)$ is pulled to the mean field $\fav$ at a fixed rate $1-\varepsilon$. 
Thus the second step acts as a {\it focusing lens}. 
The larger the coupling $\varepsilon$ is the more strongly the focusing lens acts.
Since (\ref{global1}) is a conformal contraction transformation the position of the {\it center of mass} 
$\fav$  of the system is kept invariant during the interaction. 
This is what the invariance condition (II) implies. Thanks to the invariance condition the fluctuation 
of the mean field of the maps is caused only at the first defocusing process. 
The maps in GCML evolve in time repeating this two-step process of defocusing and focusing.  
Under the battle of the two conflicting tendencies the maps eventually form an attractor 
in which fluctuations among maps is maximally suppressed by the averaging effect of the interaction 
via the mean field.

What is the condition on the coupling matrix $J_{PQ}$ for a general map model of the form (\ref{generic})
satisfies the invariance condition (II)? By summing up (\ref{generic}) over $P$ 
and using (II) we obtain a relation
\begin{eqnarray}
	\sum_{P\in\Lambda}{f(x_P)}=\sum_{Q\in \Lambda}\left( \sum_{P\in\Lambda}J_{PQ} \right)f(x_Q)\label{sumsum}.
\end{eqnarray}
At any step of the iteration this must hold. Thus the necessary and sufficient condition on the coupling $J_{PQ}$ is 
\begin{eqnarray}
	\sum_{P\in\Lambda}J_{PQ}=1, ~~~~   \forall Q\in\Lambda.   \label{unitarityGCML1}
\end{eqnarray}
From reciprocity (I) this is equivalent to 
\begin{eqnarray}
	\sum_{Q\in\Lambda}J_{PQ}=1,  ~~~~  \forall P\in\Lambda. \label{unitarityGCML2}
\end{eqnarray}

\subsection{The nearest-neighbor coupling lattice model}
This type of a model is known to show an interesting spatial cluster formation similar to the spin glass\cite{kc}.
On the other hand it is also known that the interesting two features of the GCML, i.e. the posi-nega switch and hidden coherence 
are lost.
Nonetheless it is instructive to check that this model satisfies (I) and (II) like GCML.
Let us denote the sub-lattice of $\Lambda$ as $\lambda_P$ which is a set of the nearest neighbor sites 
to $P$ including $P$ itself. 
The model is 
\begin{eqnarray}
	x^\prime_P=(1-\varepsilon)f(x_P)+\varepsilon \fav~, \label{Ising1}
\end{eqnarray}
where $\fav$ is the mean field of the maps in $\lambda_P$ at $P$, i.e.
\begin{eqnarray}
  \fav = \frac{1}{N(\lambda)} \sum_{Q\in\lambda_P}f(x_Q). \label{Ising2}
\end{eqnarray}
The coupling matrix is then
\begin{eqnarray}
J_{PQ}=\left\{
         \begin{array}{rl}
                \frac{1}{N(\lambda)} & \mbox{if $ Q \in    \lambda_P $}\\
                ~~~~0 & \mbox{if $ Q \notin \lambda_P $}  
         \end{array}
       \right.
\end{eqnarray}
where $N(\lambda)$ is the number of maps in the sub-lattice $\lambda$ in dimension $d$.
But if $Q \in \Lambda_P $, then $ P \in \Lambda_Q $.  Hence $ J_{PQ}=J_{QP} $
and the condition (I) follows.
Also we have
\begin{eqnarray}
	\sum_{Q\in\Lambda}J_{PQ}=\sum_{Q\in\lambda_P}\frac{1}{N(\lambda)}=1.
\end{eqnarray}
and (II) follows.
Note that the interaction represents a diffusion process. For instance in $d=1$ we can rewrite (\ref{Ising1}) 
and (\ref{Ising2}) as
\begin{eqnarray}
  y^\prime_P = f \left( y_P +  \frac{\varepsilon}{3} (y_{P+1}+y_{P-1}-2y_{P}) \right),   \label{laplacian}
\end{eqnarray}
and the second term  is nothing but the laplacian on the one-dimensional  lattice.
\subsection{An extend globally coupled map lattice: EGCML}
We now wish to construct the following model. Firstly it must have all to all interaction 
and satisfy the relations (I) and (II) in order to have the same defocusing and focusing dynamics as in the GCML. 
Secondly we wish to investigate the consequence of distance dependent interaction; in 
particular, we wish to know to what extent the salient features of the GCML such
as the posi-nega switch or the hidden coherence depend on the fact that the model is 
scale invariant trivially by the absence of distance scale. Do these features disappear if the interaction between the maps 
decreases with distance?  Certainly at the nearest interaction limit, we know neither the posi-nega switch nor the violation 
of the law of large numbers occurs.  Most crucial case is the interaction which decreases with distance but in a power law. 
In contrast to the exponential law this does not introduce the length scale and yet the interaction is all to all.  
Thus the model we would like to construct is an extended GCML (EGCML) in which the interaction decays with a power law
and satisfies the conditions (I) and (II).  Below we show there is such a natural extension of GCML. In this article 
we choose $d=2$ for simplicity though our construction goes through for all dimensions.

Let us denote the set of sites at distance $\rho$ from $P$ as $L_\rho(P)$. 
The symbol $L$ is the memento of $loop$ since in $d=2$ such a set is a loop (in fact a square on the lattice space) 
of radius $\rho$ around $P$. 
We denote by $N(L_\rho)$ the number of sites in $L_\rho$.
For $d=2$,  $N(L_\rho)=8 \rho ~(\rho=1,2, \cdots)$ and $N(L_0)=1$.
We denote the maximal radius of the loop realized in the lattice $\Lambda$ as $\kappa$. 
This is the perimeter of the lattice and for $d=2$, $\kappa=[(N^\frac12-1)/2]$. 
To be precise we present our result for $N^\frac12$ odd but we have checked no change of 
results occurs for $N^\frac12$ even.
In order to treat all of the sites in $\Lambda$ democratically and to minimize the finite size effect 
we impose the periodic boundary conditions. For any point $P$ on the lattice 
we calculate the effect of all maps within $\kappa$ taking into account the periodic boundary condition.    
The EGCML is then compactly described as  
\begin{eqnarray}
	x^\prime_P=(1-\varepsilon)f(x_P)+\varepsilon \favav_P. \label{TSM1}
\end{eqnarray}
where $\favav_P$ at $P$  is given by 
\begin{eqnarray}
	 \favav_P & = & \frac{1}{\kappa} \sum_{\rho=0}^{\kappa} \fav_{P,~\rho} \label{TSM2} \\   
 \fav_{P,~\rho} & = & \frac{1}{N(L_\rho)} \sum_{Q \in L_\rho(P)} f(x_Q).  \label{TSM3}
\end{eqnarray}
In words the effect of all the maps on the lattice to the map at $P$ is determined
by a double average.   
Firstly any one of the maps at an equal distance $\rho$ from $P$ must affect the map
at $P$ with the same weight. Thus they are averaged with an equal weight and this gives $\fav_{P,~ \rho}$.
Then, the effects of maps with various distances from $P$ must be accounted for.  
In the EGCML this is done again by taking an equal weight average of $\fav_{P,~ \rho}$ over $\rho$
 from zero to $\kappa$ which gives $\favav_P$ as a natural non-local extension of the diffusion process.
The first average must be an equal weight average to respect the rotational symmetry. 
The second average must be again an equal weight average. It is necessary for the invariance
condition (II) as we show below. 

The coupling matrix element $J_{PQ}$  is the weight factor for the $f(x_Q)$ to 
contribute to $x^\prime_P$ via the double average and is given by 
\begin{eqnarray}
J_{PQ}=\frac{1}{\kappa N(L_\rho)}~\propto \rho^{-(d-1)} ~~~~~~ \forall P \in \Lambda,~~ Q \in L_\rho(P).
\end{eqnarray}
Thus in EGCML any map $f(x_Q)$ at equal distance $\rho$ from $P$ contributes with an equal weight
to the mean field $\favav_P$ and the interaction decreases with an inverse power with respect to 
$\rho$ for $d=2$. 
The reciprocity (I) follows just as in the nearest neighbor case since 
$Q \in L_\rho(P) $ is equivalent to $ P \in L_\rho(Q) $.
Also the invariance condition (II) can be proved as:
\begin{eqnarray}
\sum_{Q\in\Lambda} J_{PQ} &=&\sum_{\rho=0}^\kappa \sum_{Q \in L_\rho(P)}J_{PQ}  \nonumber\\
                     &=&\sum_{\rho=0}^\kappa 
                     \frac{1}{\kappa} \left( \sum_{Q \in L_\rho(P)} \frac{1}{N(L_\rho)}
                     \right)    ~~=~~1  \label{unitarityEGCML}
\end{eqnarray}

%%%%%
\section{Phase Structure of GCML and EGCML}
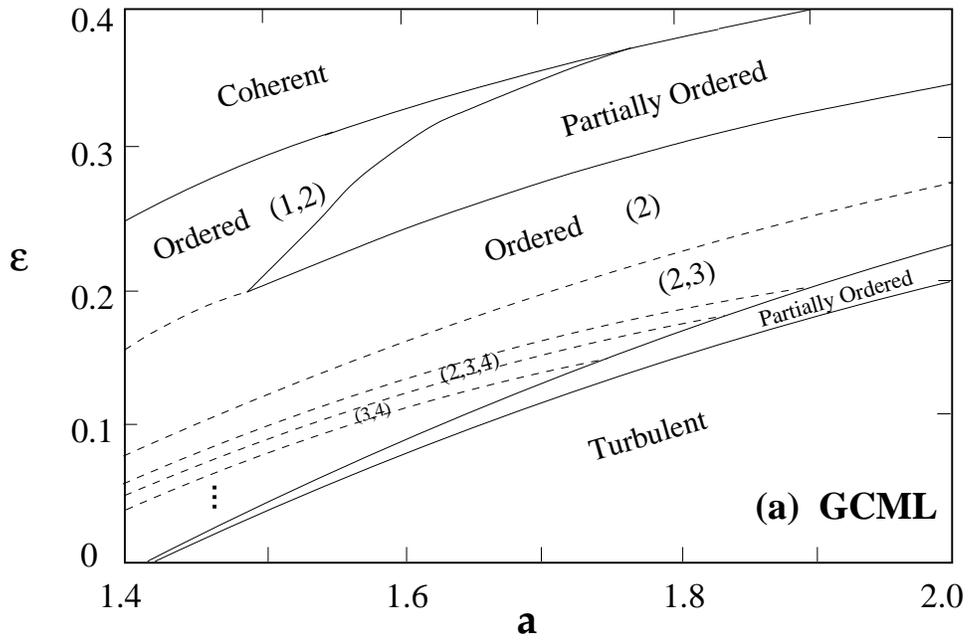
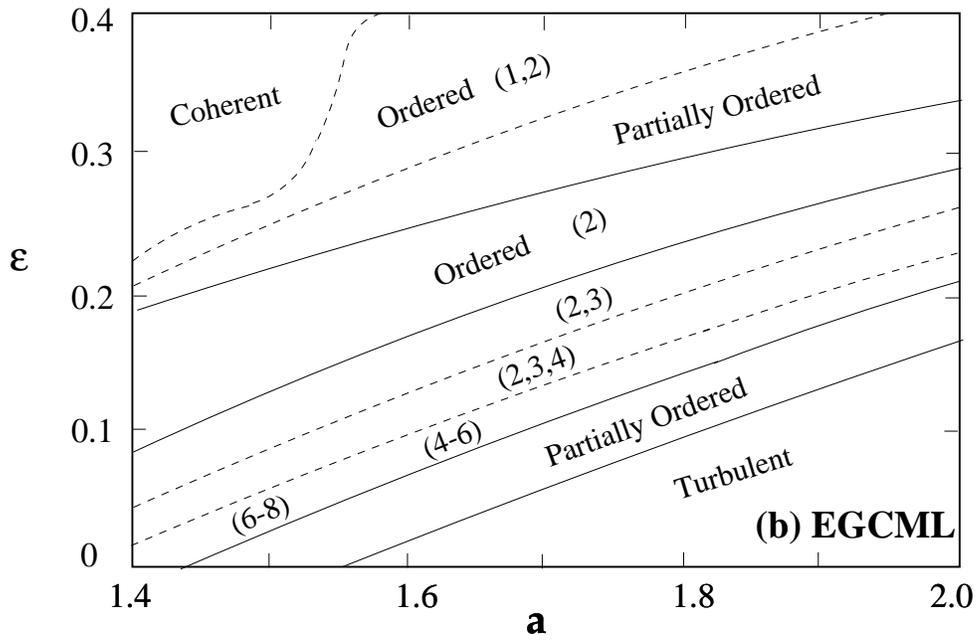
\begin{figure}
\caption{
(a) The phase diagram of GCML as determined by $N=1000$ lattice with 1000 random initial configurations  
for each $(a, \varepsilon)$. 
(b) That of EGCML.  N=$39 \times 39$ and typically 200 configurations for each $(a, \varepsilon)$.
\label{fig1}}
\end{figure}

The GCML phase diagram in the $a-\varepsilon$ plane given in \cite{ka} is reproduced in Fig.~\ref{fig1}a 
to facilitate the comparison with the EGCML phase diagram. For GCML we 
analyzed statistically the cluster composition of $N=1000$ maps in the final attractor from 1000 random initial
configurations and we verified the original phase diagram. 
The various phases are the outcome of the battle between chaotic motions of maps (the defocusing effect) 
and the interaction via the mean field forcing the maps to move coherently (the focusing lens effect). 
Since the GCML phase structure is elucidated in ref.\cite{ka} we recapitulate here only the essential features 
necessarily in the following discussion. Let us consider a line at $a = 1.8$ for definiteness which is well-beyond 
the critical value $a=1.401$ of a single logistic map to the chaos\cite{may}.  
For a sufficiently large $\varepsilon$ ($\ge 0.38)$ the maps in the final attractor are strongly bunched together
in a cluster and evolve as a unity in the same chaotic motion as that of one single logistic map. 
This is the coherent phase. For $\varepsilon$ in the range  $[0.22,0. 31]$ the interaction via the mean field can 
no longer exerts strong bunching and the final maps divide into two clusters. 
The maps in each cluster are still tightly synchronizing each other, while the two clusters mutually 
oscillate opposite in phase.   This is as a solution of a minimum fluctuation in the mean field. 
We discuss this ordered two-cluster phase extensively in Sec. IV. 

For smaller $\varepsilon$, the number of final clusters increases but the number of clusters at a given
$\varepsilon$ does not depend on the number of maps $N$. 
The typical number of clusters at various $\varepsilon$ ranges is indicated in the phase diagram.
For further smaller $\varepsilon$ ($\le 0.15$) the binding of the maps is almost resolved and 
the number of clusters is proportional to $N$. This region is the turbulent phase and we 
extensively analyze this regime in Sec. V. 
Final remark is that the boundary values of $\varepsilon$ between phases simultaneously increase (decrease) 
in order to maintain the balance between the conflicting tendencies  if the coupling $a$ is increased (decreased).  

We show in Fig.~\ref{fig1}b the corresponding phase diagram of the EGCML. 
We used $39 \times 39$ map lattice and analyzed statistically the final attractor configuration 
using typically 200 runs for each $(a, \varepsilon)$.\footnote{
We note that the EGCML run requires increasingly large computation time for large-size lattice since
the  $\favav_P$ for each map at the lattice site $P$ must be individually calculated using the inverse power law.
For $N=39^2$ EGCML, the necessary CPU time is about a hundred times of that for GCML of the same size.} 
There is a remarkable agreement between two phase diagrams except for slight shifts in the phase boundaries. 
For instance, for the EGCML at $a=1.8$, the coherent phase, the ordered two-cluster phase,  
and the turbulent regime are respectively realized at $\varepsilon \ge 0.38$, $0.22 \le \varepsilon \le 0.31$ and $\varepsilon \le 0.15$ 
to be compared with the boundaries quoted above. This agreement between two phase diagrams 
indicates strongly that EGCML is in the same universality class\cite{universality} with GCML. 

\section{The Posi-nega Switch in GCML and the Pattern Formation in EGCML}

\subsection{The Two-Cluster Regime and the $\theta$ Variable} 
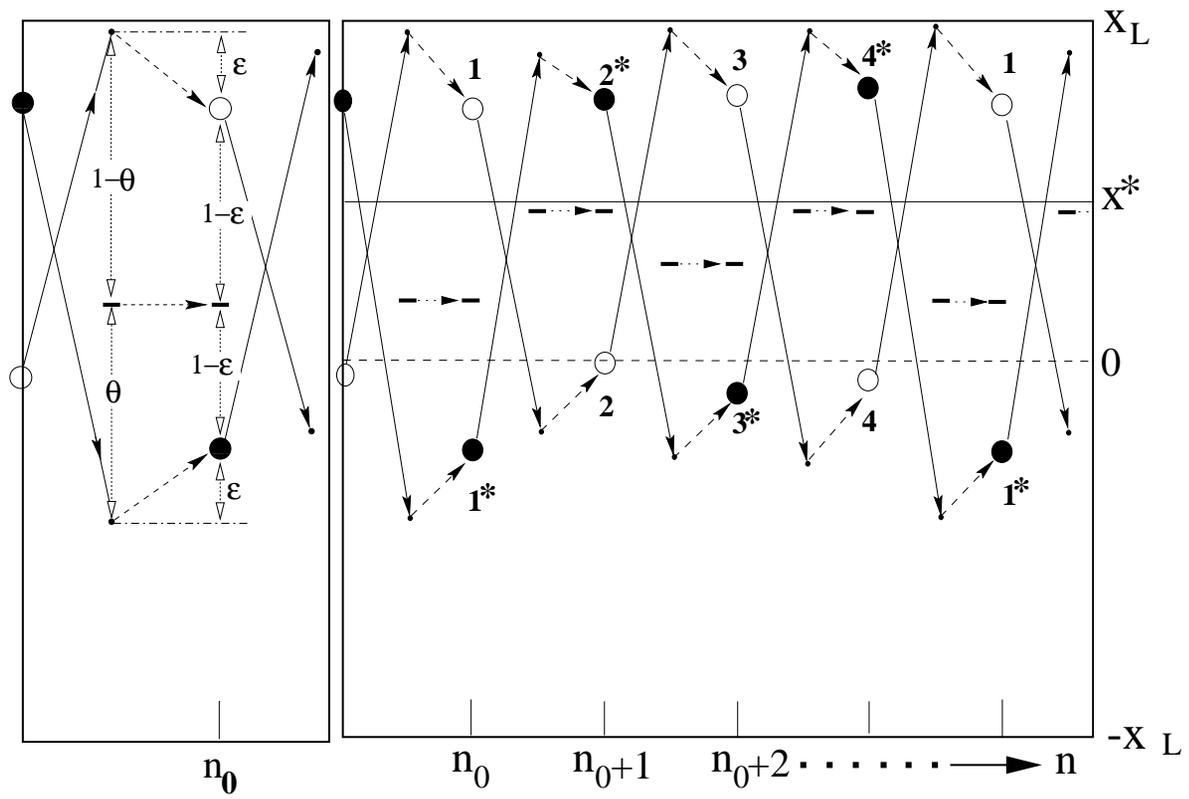
\begin{figure}
\caption{
Left box: The GCML  two-cluster dynamics. The two clusters evolve repeating 
mapping (solid arrows) and interaction (dashed arrows). 
The mean field (a short solid bar) of maps is determined by the ratio $\theta$ and $1-\theta$.  
The interaction pulls two clusters to the mean field at a rate $1-\varepsilon$ and 
hence the mean field is unchanged.
Right box: The period four attractor of two GCML clusters at $a=1.98$, $\varepsilon=0.3$. 
The black (white) circle nearer to the mean field is the majority (minority) cluster 
which involves 43 (57) percent of  maps. 
If $n_0=$ even (odd),  the white (black) circle is named as the positive cluster 
with $\theta=0.43 (0.57)$.  
\label{fig2}}
\end{figure}

The ordered two-cluster phase of GCML is realized in a band in the $a, \varepsilon$ parameter space 
with $\epsilon \in [0.2,0.23]$ at $a \approx 1.6$ and $\epsilon \in [0.28,0.33]$ at $a \approx 1.98$ 
(see Fig.~\ref{fig1}a) .
In this phase, the maps divide into two clusters after some transient steps for most of the initial values.  
The clusters oscillate around the unstable fixed point $x^*$ mutually opposite in phase and the maps 
in each of the two clusters are in tight synchronization.  
In order to distinguish the two clusters we follow the convention in \cite{ka} and name the cluster which 
takes a value above (below) $x^*$ at an even step $n$ from the first of the iteration 
as the positive (negative) cluster.
We denote the number of maps in the positive (negative) cluster as $N_{+}$($N_{-}$)
and define the population ratio parameter 
$\theta \equiv N_{+}/N$. The values of maps in the positive (negative) cluster
are uniformly  $x_{+}~(x_{-})$ if the synchronization of maps  in each cluster is really tight.
The maps are evolving in two clusters repeating the two step process of mapping and interaction 
and the dynamics of maps is reduced to that of two clusters:
\begin{equation}
x_{\pm}(n)=(1-\varepsilon)f(x_{\pm}(n-1))+\varepsilon \favn  \label{twocluster1}
\end{equation}
and the mean field of maps at time $n$ is given by
\begin{eqnarray}
 \favn &\equiv&  \frac{N_{+}}{N} f(x_{+}(n-1)) +\frac{N_{-}}{N} f(x_{-}(n-1)) \nonumber   \\  
          &=&   \theta x_{+}(n) + (1-\theta) x_{-}(n)                \nonumber         \\ 
          &=&   \langle x(n) \rangle    \label{twocluster2}  
\end{eqnarray}
where the second equality comes from the invariance condition (II) for the two-cluster configuration.  
Thus when a set of three numbers, $a$, $\varepsilon$ and $\theta$ is given the attractor of the two clusters is 
uniquely determined. 
For the discussion of the posi-nega switch below let us explain these by an explicit sample in Fig.~\ref{fig2}.
In the left box we illustrate the mechanism of Eqs. (\ref{twocluster1}) and (\ref{twocluster2})
and in the right box we show a sample attractor of two clusters realized at $a=1.98$, $\varepsilon=0.3$
and the corresponding motion of the mean field.  
The two clusters, represented by white and black circles, evolve in time 
under the iteration of mapping (solid arrows) and focusing by interaction (dashed arrows).
The mean field (a short bar) is kept invariant under the interaction as dictated by the invariance 
condition (II) and (\ref{twocluster2}). 
The first time $n_0$ in this run happened to be even. 
Hence the positive cluster is the white circle moving as $(+-+ \cdots)$ and the negative one is the black circle
moving as $(-+-\cdots)$. We see that they move in period four and mutually opposite in phase. 
The mean field is closer to the negative cluster, which means that the positive cluster is the minority cluster 
and $\theta$ must be less than half.  Numerically $\theta=0.43$ for this attractor.  
Note that it is probabilistic whether the first time $n_0$ is even or odd.  
From another initial configuration the same attractor {\it  modulo translation in time} may be equally 
realized.  If the translation is odd steps in time we will have the same figure but with $n_0$ is now odd.
The dominant cluster (black circle) must be then named as the positive cluster and $\theta=0.57$ by definition.

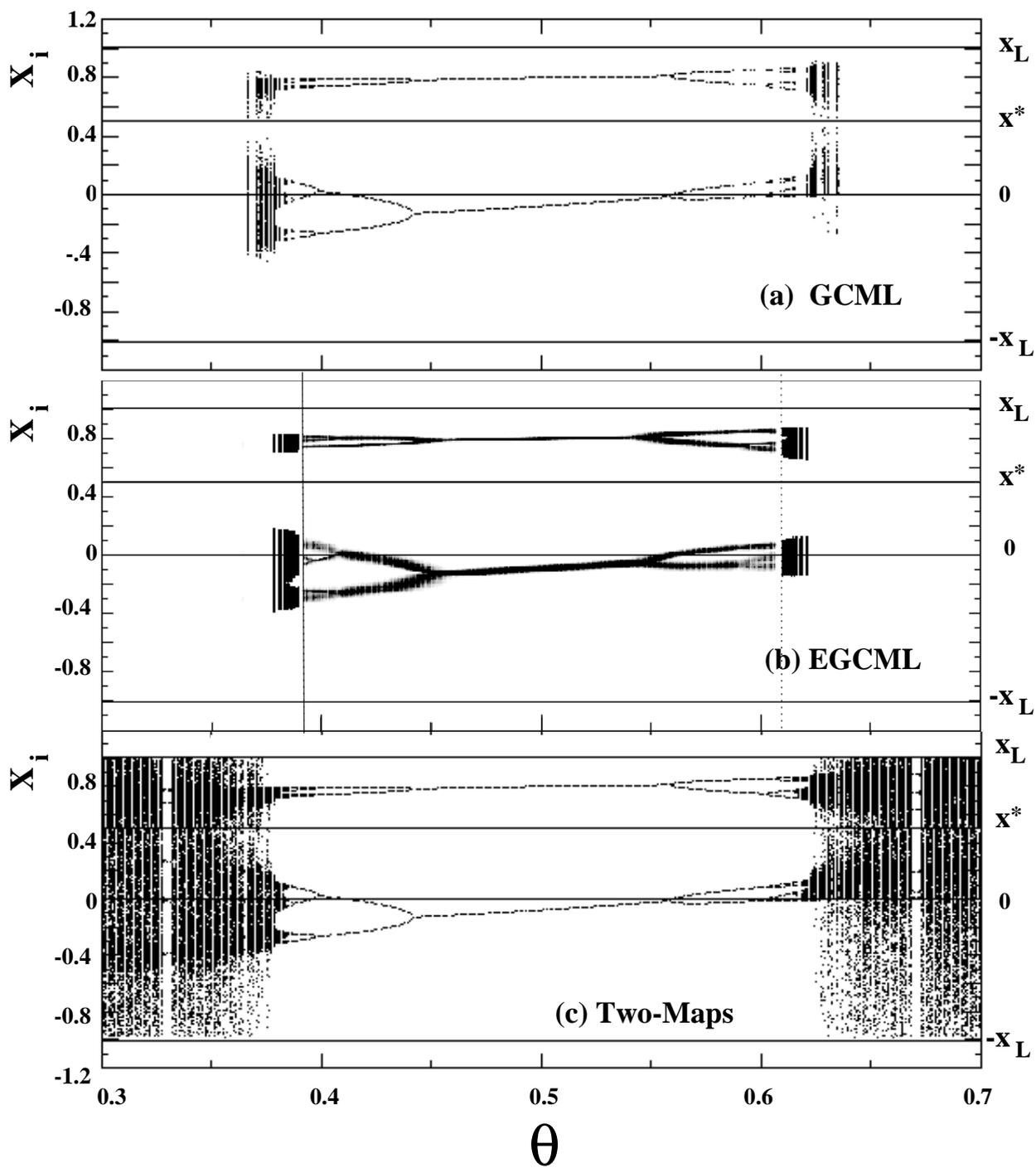
\begin{figure}
\caption{
The self-organized attractors at $a=1.98$ and $\varepsilon=0.3$.
(a) GCML  : All maps $x_i(n),~ i=1,\cdots, 1000$ from 1000 random initial configurations
are plotted at the position of the fraction of maps with $x_{i}(n) > x^*$ for iteration steps 
$n=10000,\cdots, 10512$ 
(b) EGCML :  $N=51 \times 51$ maps from 4000 initial random configurations.
The periodicity $T( \le  2^6)$ of the set of maps is measured using the judgment 
condition $\sum_{i=1}^{N} (x_{i} (n+T)-x_{j}(n))^2 < 10^{-6}$. 
Maps are sampled only for the first one cycle of the period and the points with density less 
than one percent of the maximum are removed if  they fall into a periodic attractor. 
Otherwise just as in GCML. 
The former gives the part with $\theta \in [0.39, 0.61]$ and the latter the rest.
(c) The attractor of the matrix-coupled two maps sampled for $n=10000,\cdots, 10512$
with the parameter $\theta$ varied.   
\label{fig3}}
\end{figure}

% ----------------Fork, \ref{fig3}a,  3b -----------------------
In Fig.~\ref{fig3}a and b we show the attractors of GCML and EGCML at $a=1.98$ 
and $\varepsilon=0.3$  respectively as a function of $\theta$.
We used $10^3$ runs for GCML and $4\times 10^3$ runs for EGCML. 
 In one run for GCML,  $N=1000$ maps are iterated from a random  initial configuration and 
 to account for the first transient process the values of maps during the $512$ steps after the first 
 $10^4$ iteration are plotted at the fraction of maps with $x_{i}(n) > x^*$.
For instance the run shown in Fig.~\ref{fig2} with $n_0$ even contributes four points $(1,1^*,3,3^*)$ at $\theta=0.43$
and other four points $(2,2^*,4,4^*)$ at $\theta=0.57$.
Then the run is repeated and the results of all runs are superposed in order to see the dependence of the attractor 
on the population unbalance between the clusters. 
Note that this is equivalent to plot  the maps only at the last even iteration steps at $\theta$ when 
 the results from many initial configurations are superposed.   
The GCML attractor in Fig.~\ref{fig3}a agrees with that found in ref.\cite{ka}
\footnote{We note that  the lower half attractor of GCML in \cite{ka} is 
erroneously a mirror image of the correct one.}.
When $\theta$ approaches an upper  threshold $\theta_{th}$ from below 
 (or  a lower threshold $1-\theta_{th}$ from above) the attractor undergoes successive period doubling and 
 at the thresholds the maps fall into chaos ($\theta_{th}$=$0.63$ and $1-\theta_{th}=0.37$ for $a=1.98$). 

 For  EGCML in Fig.~\ref{fig3}b the final attractor of $N=51\times51$ maps is measured in a similar way 
but in order to shorten the large computing time necessary for EGCML 
a routine is included in the iteration program which judges 
whether the maps have fallen into a periodic attractor of a certain period $T$ 
less than or equal  to $2^{6}$ or not\footnote{
In the iteration $N_{+}$ is calculated at every even step $n$ and when it stops to vary during ten successive 
steps the routine is activated at $n_0$. It calculates the judgment condition 
$\sum_i (x_i(n_0+T)-x_i(n_0))^2 < 10^{-6}$ and if this is satisfied for some $T( \le 2^6) $  the maps are judged
to have fallen into an attractor of a period $T$. If not, the routine is inactivated and the next stabilization
of $N_{+}$ is waited for.}. 
If the judgment condition is satisfied the maps are plotted at the fraction of maps with $x_(n) > x^*$ 
only during the next one cycle of the attractor and the next run is started.
This not only shortens the calculation time but also enables the statistical analysis of the frequency of the occurrence 
of the attractors and of the fluctuation of maps in a given attractor
since indefinitely duplicated plotting of the maps is avoided. 
If on the other hand the maps do not fall into an attractor all through the $10^4$ steps  
the  maps are plotted for the next $512$ steps just as in the GCML run.
In EGCML we find that the maps in a cluster synchronize more loosely than those in GCML.
The dispersion of maps in each cluster is approximately 0.1 for all $\theta$ to be compared 
with the gap size $\approx 1.0$   between the two bands of the attractor 
at $\theta \approx 0.5$. Even though the maps do not mix between the clusters in the final stable 
attractor this fluctuation hides the details of bifurcation structure with respect to $\theta$ 
in a direct superposition of all runs in terms of the bitmap graphics. 
Thus we have done a statistical  analysis of  the plot density and  removed those points with density 
less than one percent of the maximum density from the diagram. 
This procedure is justified because we truncated each run just after one cycle of the periodic attractor 
so that each run contributes to the points on the attractor diagram with an equal weight.  
It was a remarkable instance in this work when we observed the same bifurcation diagram 
with GCML turned out in the central region 
of the EGCML attractor ($\theta \in [0.38, 0.62]$)  as is shown in Fig.~\ref{fig3}b.
 In EGCML the synchronization of maps in a cluster becomes looser than that of GCML but still the dynamics of final 
clusters is common for both GCML and EGCML and controlled by $\theta$.  
We have already seen above that EGCML is endowed with a phase diagram remarkably similar to that of GCML
and now we find that the attractors in both models in the ordered two-cluster regime are also similar.   

The posi-nega switch in GCML uses the ability of $\theta$ to control the attractor.
But before delving into the detail discussion of the posi-nega switch let us examine to what extent the reduction 
of dynamics of maps is realized.
If the reduction is really effective down to two degrees of freedom, 
all $N_{+}$($N_{-}$) maps in the positive (negative) cluster 
should have a single value  $x_{+}(n)$ ($x_{-}(n)$).  
Then Eqs. (\ref{global1}) and (\ref{global2}) should reduce to (\ref{twocluster1}) 
and (\ref{twocluster2}) or equivalently to a matrix coupled two-map model
\begin{eqnarray}
   x_{+}^\prime &=& (1-\varepsilon_{-}) f(x_{+}) + \varepsilon_{-} f(x_{-}), \label{matrix}\\
   x_{-}^\prime &=& (1-\varepsilon_{+}) f(x_{-}) + \varepsilon_{+} f(x_{+}), \nonumber
\end{eqnarray} 
with $\varepsilon_{+}= \varepsilon\theta$ and $\varepsilon_{-}=\varepsilon(1-\theta)$.
From the remarkable agreement between GCML attractor (Fig.~\ref{fig3}a) 
and two-map model attractor (Fig.~\ref{fig3}c) we see that in the two-cluster phase of GCML 
the dynamics of the $N$ maps reduces to that of two maps
 for all $\theta$ between the thresholds.  

%revision re. criticism (4)
In EGCML the coupling $J_{PQ}$ is not a single constant but it depends on the distance between $P$and $Q$.
The interaction acts on a map at a site $P$ via the position dependent mean field $\favav_P$ which is a double 
sum given by (\ref{TSM2}) and (\ref{TSM3}) contrary to a single sum $\fav$ given by (\ref{global2}) in GCML. 
Therefore the reduction of EGCML to two maps we find in Fig. 3b requires an explanation. 
What is important here is not the uniformness of the couplings but the scale invariance of them.
In both GCML and EGCML the interaction has no cutoff scale and hence it prevails over the lattice 
and the maps are forced to synchronize among themselves due to the averaging effect of the 
interaction via the mean field. Thus the possible difference of EGCML from GCML comes solely from 
the variation of the site dependent $\favav_{P}$ around $\fav$. 
If the variation is small both models will show similar attractors.
EGCML is constructed in such a way  to assure this small variance 
when the interaction forces the maps to synchronize.

First let us consider a simple case where the nonlinear parameter $a$ is so small 
that the chaotic randomness in each map cannot overcome the averaging effect. Then after iterations 
all the maps on the lattice will synchronize and take approximately the same values 
with only a small variance. If we ignore this variance, the value $f(x_P)$ common to all $P$
becomes a multiplicative factor to the sum for the mean field and $\favav_{P}  = \fav$ follows 
from the invariance conditions (\ref{unitarityGCML1}) for GCML and (\ref{unitarityEGCML}) for EGCML.  
In GCML the variance is negligible but in EGCML the weaker 
synchronization causes a variance typically to the order $10^{-2}$ in $\favav_{P}$ 
and the EGCML attractor turns out reflecting this variance. But otherwise EGCML behaves just like GCML. 

Now in the two-cluster phase at higher $a$  the maps cannot be bunched together in a single 
cluster and divide into two clusters, A and B.
The maps in each of the clusters  take approximately the same value and synchronize
among themselves. Hence we may write 
\begin{eqnarray}
	x_i &=& X_A  + q_i      ~~~( i \in A)  \nonumber \\ 
	x_j &=& X_B  + q_j      ~~~( j \in B)  \nonumber
\end{eqnarray}
Here $ X_A$ ($X_B$) is the collective coordinate for A (B) (the average of maps taken over A (B) only)
and the $q_i $ ($q_j$) represents small fluctuation around the $X_A$ ($X_B$).
Now let us analyze $\favav_{P}$ for EGCML. 
An important observation is that the double sum for $\favav_{P}$ can be written as a 
sum of two double sums, one only over cluster A and the other only over cluster B. 
This decomposition allows one to apply the above argument to each of the clusters 
and the $f(X_A)$ ($f(X_B)$) becomes a multiplicative factor to the double sum 
for cluster $A$ ($B$). 
As for the remaining double sum of couplings, some more consideration is necessary. 
In EGCML a `polarized' distribution of maps is in principle allowed 
in which, for instance,  the maps in the cluster $A$ ($B$) accumulate around 
a given point $P$ and maps in cluster $B$ ($A$) accumulate far from $P$. 
But EGCML has no interaction to induce such a polarization 
and the randomness in each map also acts to break the polarization.  
We have numerically checked the dominance of unpolarized distributions. 
Therefore the remaining double sum for EGCML again counts the number of maps 
in each cluster just in the same way the single sum for GCML does in (\ref{twocluster2}). 
Thus we have also in EGCML 
\begin{eqnarray}
    \favav_{P}  \approx \frac{ N_A}{N}f(X_A)  + \frac{N_B}{N}f(X_B)
\end{eqnarray}   
Here the approximation takes into account both the polarization in fluctuation 
and the small variance of maps in each cluster due to approximate synchronization.  
Thus the collective coordinates $X_A$ and $X_B$ of the clusters is described by the matrix 
coupled two-map model and the maps in each of the clusters fluctuate around the collective coordinates. 
This explains the approximate reduction of EGCML to  two maps.

% In EGCML the synchronization of maps in each cluster is looser than in GCML and the EGCML attractor 
% turns out as bands in Fig.~\ref{fig3}b rather than sharp lines. 
% However the maps in each cluster are in precise phase 
%synchronization\cite{fns,fs,pikovskya} so that no mixing occurs between positive and negative clusters. 
% Thus dynamics of  maps in EGCML should be described by that of two mean fields of the respective clusters
%  with fluctuations around them. The agreement of the bifurcation structure of EGCML attractor with 
% that of two maps verifies the dynamical reduction in this way.

\subsection{A resolution of the posi-nega switch mechanism}

The posi-nega switch found in GCML\cite{ka} is based on the ability of $\theta$ as the control parameter 
of the GCLM attractor.  If  maps are successively transported from one cluster to the other by input 
($x_i \longrightarrow x_i \pm \delta$) the population unbalance can be enhanced and the attractors undergo 
successive bifurcation. 
When $\theta$ reaches the threshold the whole maps go into a grand chaotic motion almost bunched 
together in a unit. See Fig.~\ref{fig4} for a typical sample of this transition.  
This state induced by an excess of the last one map
is unstable as we see in Fig.~\ref{fig3}a that there is no self-organized stable attractor 
above $\theta_{th}$ (or below $1 - \theta_{th}$).
Thus after some transient time the maps again form the allowed stable 
configuration, that is,  tightly bound two clusters
mutually oscillating opposite in phase.  
Just as before let us again label the new cluster as positive if it  takes a value above
$x^*$ at even step $n$ (counting from the start of iteration) and the other as negative. 
An amazing finding in \cite{ka} is that the composition of the new two clusters is limited only in two ways.
(1) All the maps involved in the positive cluster before the chaotic transient are now involved in the negative cluster.
(2) All the maps in the old positive cluster are again in the new positive cluster.
The case (1) is the posi-nega switch.  In both cases the maps behave as if they keep a {\it memory} of their former 
clusters.
But how is it possible?  The maps pass the chaotic transient process where many two-cluster states
as well as many other channels are open.  What is the mechanism that protects the maps from mixing? 
   
Before answering this old puzzle let us point out some trivial aspects of the posi-nega switch. 
First there is no essential difference between case (1) and (2). How long the chaotic transient process persists
and in particular whether the duration time is even or odd is (almost) probabilistic. 
It is of course deterministic but highly dependent on the map values at the onset of the transient process. 
Whether a cluster after the transition is positive or negative depend on 
the parity (even or odd) of the duration time by definition and there is no sense to distinguish (1) and (2)
when the parity is indefinite. 
Second we should note that in the two-cluster phase of GCML with fixed $a$ and $\varepsilon$ 
the self-organized attractor   
is determined solely by the population unbalance between the two clusters.
In case (1)  $\theta \longrightarrow 1- \theta$ and in case (2) $\theta \longrightarrow \theta$.
In either case,  the population {\it unbalance} is the same before and after the transition and hence the attractor before 
and after must be the same. To be precise, to put the attractor before the transition upon that after the transition 
 we need a temporal translation of odd steps in case (1) and even steps in case (2).  Modulo this difference 
 the attractor before and after the transition are the same in either case. 
 There is no chance of appearance of a new attractor 
 {\it provided that}  the population {\it unbalance} before and after the transition is the same
 \footnote{In ref.\cite{ka} this point that the maps come back to the same attractor is not noted.}. 
 Therefore the real puzzle boils down to a single question. What is the mechanism which keeps the population unbalance 
 between two clusters unchanged before and after the chaotic transient process?
We solve this puzzle below in two steps. First we discuss how the chaotic transient process starts and ends. 
This provides the temporal boundary conditions of the transient process. 
Then we show there is a very simple mechanism which transfers the population unbalance rate 
through the transient process.

\subsubsection{The transition at the threshold} 
%--------------------mapping mechanism --------------
The logistic map $f(x)$ copies the region $[-x_L, x_L]$ onto itself twice-fold as follows\footnote{
For the map $f(x)=1-ax^2$ the limits are $\pm x_L=\pm(1+\sqrt{1+4a})/2a$ and 
the unstable fix point is $x^*=(-1+\sqrt{1+4a})/2a$ but the argument here is independent of 
the particular parametrization of the map. All we need is the universal folding nature of the map.}.
\begin{eqnarray*}
   [-x_L, x_L] \equiv I+II+III  \nonumber \\
   I &\equiv& [-x_L, 0] \stackrel{f}{\longrightarrow}  I+II+III \nonumber \\
   II&\equiv& [0, x^*] \stackrel{f}{\longrightarrow}   III              \\     %\label{logisticmap}
   III&\equiv&[x^*, x_L] \stackrel{f}{\longrightarrow} II+I ~.  \nonumber \\ 
\end{eqnarray*}
The first two mappings are monotonously increasing while the last is monotonously decreasing.  
See Fig.~\ref{fig5}.
Below the threshold the two clusters are moving mutually opposite in phase around $x^*$ so that one must be in region 
III and the other in region I or II. 
When the unbalance is increased by sending the maps of the minority cluster to the majority cluster 
the mean field comes closer to the majority cluster and oscillates in phase with it. 
The interaction conformally contracts the maps to the mean field at every iteration step, and the 
minority cluster is pulled to the mean field more strongly at this highly  unbalanced situation. 
In  a few iterations after the unbalance exceeds the threshold by the last input, 
the non-dominant cluster crosses the boundary $x^*$. 
This is the very time of the start of grand chaotic transient motion of maps.
At the next step of iteration all maps must move into the same direction. 
The synchronization opposite in phase ceases and the maps start almost coherent evolution. 
The mean field evolves together in the middle of them. The focusing effect of the interaction now works only 
to bunch the maps close together around the mean field.
The high nonlinearity of each map then becomes manifest and all maps move around the three regions 
bunched together in a pseudo-cluster with a slight variance around their mean field.
In the typical sample run shown in Fig.~\ref{fig4} the mean field (depicted by a short sold bar) 
oscillates closely to the dominant cluster (connected by the solid line) for several steps.  
At time $n_1$  the minority cluster (the dashed line)  is pulled down across the $x^*$ and 
at the next step both clusters are mapped upwards together. Then the grand chaotic motion starts at $n_2$.
The pseudo-cluster is formed by the external input and is unstable. 
To bunch it stably the coupling must be larger; it must have the value for the coherent region (e.g., $\varepsilon > 
0.43$ for $a=1.98$). The tiny variance between maps in the pseudo-cluster is 
soon magnified and the transient pseudo-cluster is broken to the self-organized stable two clusters again.  
In the following we use the fact that the maps starts and exits the transient process with small variance
among them. 

\subsubsection{The population conservation mechanism during the chaotic process}
%--------------------  relative co-ordinates -------------
Let us define the relative coordinates $\tilde{x}_i$ with respect to the mean field 
\begin{eqnarray}
	\tilde{x}_i(n) &\equiv& x_i(n)  - \langle x_i(n)\rangle  , \\
	\langle x_i(n)\rangle &\equiv& \frac1N \sum_{i=1}^N x_i(n).  \label{xrel}
\end{eqnarray}
Since all of the maps are now close to the mean field we may expand
$f(x_i(n))$ as
\begin{eqnarray}
	 f(x_i(n)) = f(\langle x_i(n)\rangle  ) + 
 	\tilde{x}_i(n)\frac{df}{dx} \Biggl| _{{}_{\langle x_i(n)\rangle  }}   \Biggr.
 	+O(max(\tilde{x}_i^2)). \label{Taylorexpansion}
\end{eqnarray}
Averaging this over $i$ using $\sum_{i=1}^N \tilde{x}_i(n)=0$ we obtain a relation 
\begin{eqnarray}
	  \langle f(x_i(n))\rangle   \equiv \frac1N\sum_{i=1}^N f(x_i(n))=f(\langle x_i(n)\rangle)
   	+ O(max(\tilde{x}_i^2)). \label{averagerelation}
\end{eqnarray}
But the interaction does not affect the mean field (the condition (II)). So we obtain 
\begin{eqnarray} 
  	\langle x(n+1)\rangle = \langle f(x_i(n))\rangle = f(\langle x_i(n)\rangle  ) 
  	+ O(max(\tilde{x}_i^2)).
\end{eqnarray}
Therefore the evolution equation of the relative coordinates is given by
\begin{eqnarray}
 \tilde{x}_i(n+1) &\equiv& x_i(n+1) -\langle x_i(n+1)\rangle   \\ \nonumber 
&=&\Bigg((1-\varepsilon) f(x_i(n))
+ \varepsilon\langle f(x_i(n))\rangle\Bigg) - \Bigg( f(\langle x_i(n)\rangle  ) 
+ O(max(\tilde{x}_i^2))  \Bigg)   \\ \nonumber
&=&   (1-\varepsilon)\Bigg(f(x_i(n))-f(\langle x_i(n)\rangle  ) \Bigg)  + O(max(\tilde{x}_i^2))  \\ \nonumber
&=&(1-\varepsilon)  \frac{df}{dx}
\Biggl|_{\langle x_i(n)\rangle  } \Biggr.\tilde{x}_i(n)  + O(max(\tilde{x}_i^2)). \label{mechanism}\nonumber 
\end{eqnarray}
We used (\ref{averagerelation}) once more in order to derive the third line. 

The last line is crucial. The factor $(1-\varepsilon) df/dx\left|_{\langle x_i(n)\rangle}\right.$ is 
common to all maps
and in particular it acts with a common sign.
Thus at every step in the chaotic transient process the maps separated by the mean field 
never mix each other.
This is the mechanism which keeps the population unbalance unchanged before 
and after the chaotic transient process.
The sign of the factor at $n$ depends on the mean field value $\langle x_i(n)\rangle$. 
Hence case (1) and (2) may occur equally.
We note that  the invariance condition (II) is used in proving the mechanism.
This condition is really important to explore the model which shares the property of GCML.  

In the above resolution we examined the case in which the maps are confined in a pseudo-cluster 
and their variance around the mean field is kept small due to synchronization. 
Numerically we can trace the motion of the pseudo-cluster. 
Almost always the maps moves in a pseudo-cluster around the mean field with squared variance 
less than $10^{-4}$. But sometimes it intermittently splits into two or more tight clusters 
around the mean field. This does not invalidate above resolution.  
The danger of the mixing of members between clusters occurs only when 
they come close each other. 
Our mechanism gives a guarantee that no mixing occurs even in such emergence.

% ---------- numerical work
\subsection{Numerical tests of the GCML posi-nega switch mechanism}
In Fig.~\ref{fig6}a we show a posi-nega switch in $N=60$ GCML ($a=1.98, \varepsilon=0.30$) . 
In the upper (lower) diagram the maps at the even (odd) iteration steps are plotted and connected 
by lines. A similar figure was given in the first report of the GCML switch\cite{ka} but the evolution of maps 
only at the even iteration steps was presented. By such a diagram only it looks as though the attractor 
undergoes a transition to a new periodic attractor after the posi-nega switch. 
But as we noted above,  the attractor before and after the switch is the same modulo temporal transport 
of odd steps. 
We can see clearly in Fig.~\ref{fig6}a the attractor on the whole, the even and odd iteration steps together, is precisely 
the same before and after the chaotic transition. As indicated by the arrows the switch from positive to negative is trivially 
a change in the count (even or odd) of iteration steps. 
What is not trivial is that the memory of the composition of two clusters is kept 
during the posi-nega switch. 
In Fig.~\ref{fig6}b we present a posi-nega switch event of $N=50$ GCML at even iteration steps 
only as a three-dimensional landscape plot.  
The peaks (valleys) before the chaotic transient process become the valleys (peaks) which is the
posi-nega switch. In the transient process the maps form a surface which is chaotically oscillating 
but at any instance the segment of the surface looks remarkably horizontal. 
This shows that in the transient process the maps are in quasi-coherent motion with only a tiny variance.
 In Fig.~\ref{fig6}c we show a {\it print circuit pattern} of  the same switch event only at even iteration steps   
distinguishing the maps with respect to the mean field value. 
Those maps with ($x_{i}(2n)>\langle x(2n) \rangle$) are shown with white squares and 
the others by black squares. Just as we proved above,  the pattern clearly shows that there is no mixing 
of maps across the mean field all through the switch event.  It can be also seen in Fig.~\ref{fig6}c that 
the transient process ceases when the last transferred map ($n=12$) comes back to its home cluster. 
This is a typical way of the end of GCML switch. 
This completes our resolution of the posi-nega switch mechanism.

\subsection{The spatial cluster formation in the EGCML switch}

The EGCML has almost the same phase diagram with GCML as we showed in Sec. III and 
the attractor of EGCML in the two-cluster phase has the same bifurcation tree structure with that of GCML.
But as we showed in Fig.~\ref{fig3}b the attractor of  EGCML turns out with some variance. 
That is, the dynamics of clusters in EGCML is reduced to that of matrix-coupled of two maps 
just as in GCML but the maps in each EGCML cluster are in  phase-synchronization rather 
than in tight synchronization in contrast to GCML.  
The all to all interaction with no cutoff scale in EGCML is effective enough to 
maintain the reduction to two-cluster dynamics for any $\theta$ between $[1-\theta_{th}, \theta_{th}]$. 

In the numerical simulation of EGCML we find that the two clusters of EGCML undergo a switch process 
by inputs which is similar to that in GCML. A typical sample of the EGCML switch is shown in 
Fig.~\ref{fig7}. 
But let us first list the general differences of EGCML switch from GCML switch below 
and add description particular to Fig.~\ref{fig7} afterwards.

\noindent
(1) {\it Slower Reaction}~: By an injection of a pulse we transfer a map in the minority cluster to the majority cluster. 
In GCML the reaction of the attractor to this input is quite rapid and the attractor shifts to the new 
movement only after a few  steps of the iteration. On the other hand in EGCML the attractor reacts to the input 
gradually in several iteration steps.
This has a simple reason. The synchronization of maps in an EGCML cluster is looser. 
When a map is transferred there occurs a negotiation among maps in each cluster to reform 
the clusters and then the two clusters mutually move 
to the new attractor. 
This is similar to the two step phase-synchronization found in the globally coupled flows\cite{fns,fs}.

\noindent  
(2) {\it Self-Organized Transition}~: In EGCML the effect of one pulse depends on the unbalance between 
clusters at the time of injection.
If the unbalance is not high just a slow reaction of the attractor described above occurs. 
If the unbalance is high the map transportation by an input is followed by a self-movement of another map between
clusters in the same direction. This occurs after a few tens of steps for the EGCML of the size $N \approx 10^3$.
If the unbalance is close to the threshold determined by the above statistical analysis (Fig.~\ref{fig3}b) the 
transportation of a single map triggers a cascade. 
   The last injected pulse induces follower maps to move from a cluster to the other cluster 
   in the same direction and in some 100 iteration steps the move grows up to a series of 
  avalanches at various scales. 
  The corresponding change of the spatial distribution of positive and negative clusters is similar 
  to a percolation transition. Under this process no more external input is given and the maps 
  are themselves forming a metastable state. Both the existence of a threshold and the formation  
  of clusters at various scales suggest a remarkable similarity to the self-organized criticality,  e.g. \cite{Bak}, 
  but further investigation on a larger size lattice, in particular the measurement of the critical 
  exponent, is required to verify the relation.  

\noindent
(3) {\it The Mixing}~: In EGCML the two-cluster attractor after the chaotic transient process is for most 
cases approximately the same with that before. The change of  the population unbalance is only 
a few percent in most of the runs.
But  there occurs swapping of maps between clusters and the composition of the clusters in terms of maps 
is often different before and after the EGCML switch.  
The posi-nega switch in GCML is realized by the fact that no mixing of maps across the mean field 
occur during the chaotic transient process.  In EGCML the mixing does occur since each map is 
under the influence of position-dependent $\favav_P$. Thus the EGCML switch is a posi-nega switch
as a cluster process but not as the map process.  In view of the brain dynamics this difference may be immaterial
since what is important is the activity of whole neurons rather than the identity of individual neurons.

\noindent
(4) {\it Self-Organized Spatial Clusters in EGCML}~: Before the switch the maps in the positive cluster and 
those in the negative cluster are almost randomly distributed on the lattice. 
The transition to chaotic transient process goes like a percolation
transition. When the self-organized mass movement starts,  the positive maps form several large-size spatial 
clusters and so do the negative maps. Since these spatial clusters are formed by the long-range interaction
the formation of them is more remarkable in the lattice of larger size.
The spatial clusters can be best traced in the chaotic process by distinguishing the maps
by the mean field of whole maps.  At short time scale often we observe a posi-nega switch of spatial patterns. 
But due to mixing the spatial clustering pattern changes at longer time scale.

%------------EGCML -------- Fig 7.-----------------------------
Now we explain  Fig.~\ref{fig7}a which is a typical EGCML switch event  sampled 
in $25 \times 25 $ lattice ($a=1.9, \varepsilon=0.3$).

\noindent
{\it  Preparation:} The maps are iterated from random start and a few tens of pulses are 
injected during the first 1500 iteration steps until  the two synchronizing clusters exhibit period 32 
oscillation mutually opposite in phase.  

\noindent
{\it Start of Chaotic Process:} At $n=1630$ we injected ten pulses simultaneously 
to randomly chosen maps on the lattice.
This was necessary to show the whole process in one figure. 
Had we injected the pulses one by one with say a hundred steps
 in between we had been able to see the accelerated cascade transition 
 as described in (2) above. But still we can see in Fig.~\ref{fig7}a that the process 
 after $n \approx 1700$ is a  self-organized cascading process. 

\noindent
{\it  The Posi-nega Switch between Clusters:} At $n \approx 2400$  the transient process 
ceased and a periodic attractor of two clusters appeared.
As indicated by arrows we see clearly that it is  the same attractor as before.
The change of the unbalance of population between the two clusters is less than one percent in this switch.

\noindent
{\it The Spatial Clusters:} The snap-shot patterns of the maps at the even iteration steps 
are shown in the boxes below the evolution plot. In the upper boxes we distinguish the maps with 
respect to the mean field while in the lower boxes we used the unstable fixed point $x^*$.
In either way we can see that spatial clusters are formed by the self-interaction of maps during 
the chaotic transient process. 
We see clearly that the maps undergo a posi-nega switch among themselves between snap-shots 4-7 
in the upper boxes. On the other hand in the lower boxes the pattern becomes sometimes 
totally black and sometimes totally white. 
Hence, we see that the distinction by the mean field is a better way to sense the spatial cluster 
formation on the lattice. A distinction by the mean field is sensitive to the fluctuation around the collective 
coordinate. The clustering  pattern extracted by this method is the long-range fluctuation 
dynamics of maps on the lattice.

This long-range interaction dynamics can be seen more remarkably on a larger lattice 
which allows the formation of the spatial pattern of longer wavelength. In Fig.~\ref{fig7}b we 
show the samples of  amazing spatial patterns formed during the chaotic transient process 
in the run of $61 \times 61$ lattice ($a=1.98, \varepsilon=0.3$). The snap-shots  were sequentially 
taken at every four iteration steps.
   We observe that a small seed of fluctuation tends to grow into a circular dense pattern by the  
   chaotic diffusion process on the lattice in accord with the isotropic interaction in EGCML.  
   It is interesting to note the similarity of the patterns observed on the lattice to the galaxies in the universe 
   and in particular the appearance of patterns like globular nebula. The EGCML may be used as a toy model 
   for the formation of density fluctuations in a nonlinear diffusion process, though further study is required 
   to test this speculation.

\section{The turbulent regime of GCML and EGCML}
The turbulent regime is generically the region of small $\varepsilon$ and known by the so-called hidden 
coherence phenomenon.  Generally it is understood that in this region  
maps evolve almost randomly  at small lattice size $N$ and that  there occurs the hidden coherence at large $N$.
But actually the turbulent regime has a subtler feature.
 We are going to show that it is a regime  where the grand periodic or quasi-periodic motion of the whole 
 maps is in action  almost everywhere. But first let us quote briefly previous observations in the literature
 in three items.

(i) When GCML is iterated from a random configuration with a  small coupling ($\varepsilon < 0.2$) and a
large  nonlinear parameter ($a > 1.6$) it reaches a final state which is an ensemble
 of maps and tiny clusters, each moving chaotically due to  high nonlinearity of the map. 
 (We show below that they are under certain coherence almost everywhere in the turbulent regime
even though each element looks in an independent random motion.) 
The number of elements increases proportionally to the number of whole maps 
in sharp contrast to the ordered regime where the dynamics of maps reduces to that of a few clusters\cite{ka}.

(ii) It was found in the statistical analysis of the mean field fluctuations that there emerges 
certain coherence between elements for large lattice size $N$\cite{kb}. 
(We show below some reservation is necessary to this statement.)
Let us denote the mean field of the $N$ maps on the lattice $\Lambda$ at time $n$ as
\begin{equation}
 h_n \equiv \frac{1}{N}\sum_{i \in \Lambda} x_i(n) 
 =\frac{1}{N}\sum_{i \in \Lambda} f(x_i(n-1)). 
\end{equation}
If $x_i(n)~(i=1,\cdots N)$ are really $N$ independent random variables with a common distribution,  
the mean square deviation (MSD) of the fluctuation ($\delta h^2 = {\langle h^2 \rangle}-{{\langle h \rangle}}^2$) 
should decrease proportionally to $1/N$ by the law of large numbers (LLN) and 
the distribution of the $h_n$ must be a gaussian for sufficiently large $N$
by the central limit theorem (CLT).  
However the time series analysis of the mean field fluctuation shows  that there is a threshold $N_{\delta}$
 in $N$ (depending on both $a$ and $\varepsilon$) above which MSD ceases to shrink 
 even though the distributions remains in a  gaussian shape. 
Even though the elements (maps and tiny clusters) look evolving independently each other
the violation of LLN must reflect some hidden coherence between the elements.  
In fact an introduction of a very small independent noise term in each map restores LLN\cite{kb,pikovskyb}.

(iii) This violation of LLN observed in the time averaged distribution of $h_n$ reflects 
that the probability distribution $\rho(x)$ that a map takes a value $x$ 
explicitly depends on time. A noise intensity analysis of ensembles successfully proves LLN\cite{pikovskyb}.
 If LLN should hold in the time average, $\rho(x)$ would have to be the fixed point distribution 
 of the Frobenius-Perron(FP) evolution equation\cite{Ruelle}. 
It has been shown that the fixed point distribution is unstable due to the periodic windows inherent 
to the chaotic logistic map\cite{kb,pikovskyb}. The coherence manifests itself in the mutual 
information\cite{kb}. On the other hand the temporal correlation function 
similar to the Edwards-Anderson order parameter for the spin glass\cite{spinglass}  
decays to zero exponentially.  Thus it may not be due to freezing between two elements\cite{kb}.

All these are important observations for the turbulent regime of GCML. 
But this is not the whole story. Especially one should not simply conclude by the above list  
that the turbulent regime of GCML (and also EGCML)
is a regime with complete randomness at small $N$ and with the hidden coherence for larger $N$.  
The MSD and distributions previously reported \cite{kb} 
are only for the limited regions of the turbulent regime where the mean field 
distributions looks semi-gaussian and the violation of LLN for large $N$ is remarkable.
But there do exist regions in the turbulent regime where the coherence is suppressed 
strongly and the MSD obeys LLN in a good approximation. 
Oppositely there are also large regions where the final attractors 
are two or three clusters which evolve stably (until the end of $10^6$ iterations) 
in period three; a grand periodicity in the turbulent regime.
Furthermore even in a large part of the \lq hidden coherence\rq~ region 
the distribution is not really gaussian; it is only semi-gaussian (trapezoidal) or has visible peaks
on top of semi-gaussian distribution. 
We can detect the remains of the periodic coherent motion by an analysis of the decay of  a
temporal correlation function for these regions.  
The turbulent regime is actually the region which is full of periodicity remnants and 
the hidden coherence is realized in between the genuinely turbulent regions and 
manifestly periodic regions. The EGCML shares the same features with GCML.
In the following we show these facts by an extensive analysis over the whole turbulent regime.

\subsection{An extensive statistical analysis of the turbulent regime of GCML and EGCML}
We measure the distribution of $h_n$ and its MSD at every point on 
the grid on the $N-\varepsilon$ plane in the turbulent regime at a fixed value of 
the nonlinear parameter $a$.  
Here we choose $a=1.9$ as a canonical value but our observations below generally hold 
for other values of $a$ larger than $1.6$. At $a=1.90$ the turbulent regime is 
from $\varepsilon=0$ to $0.15$ for GCML and from $\varepsilon=0$ to $0.12$ 
for EGCML (see Fig.~\ref{fig1} a and b).  Thus we set a $25 \times 61$ $N-\varepsilon$ grid
where the lattice size $N \equiv k^2$ varies from $k=3$ to $51$ with an increment 
of $k$ two\footnote{This increment of $N$ is necessary for two-dimensional EGCML
with the periodic boundary condition. For GCML we keep the increment the same
 for the sake of comparison. }   
and the coupling $\varepsilon$ varies from $0$ to $0.12$ with an increment $0.002$.
At every point on the $N-\varepsilon$ grid with $a$ fixed at $1.90$, 
the maps are iterated from a random start and the mean field $h_n$ is measured during  
the time interval between $n=10^4$ and $10^5$. 
Thus the result for one run at $a=1.90$ is a $25 \times 61$ table of distributions and 
each distribution is a histogram of $9 \times 10^4~$  $h_n$. 
We have repeated the run several times for the same $a$ and verified that 
the general result is independent from the change of the initial configuration of 
randomly produced maps. 

We have also checked in two ways that our $N-\varepsilon$ grid is sufficiently fine
to see the general feature of the turbulent regime.
Firstly we randomly chose more than a hundred squares of the grid and subdivided them.
Using this finer grid we have verified that either the variation of $\varepsilon$ or $N$ within each square 
does not cause any significant change of the distribution (a local check).
Secondly the compiled $1525 (=25\times 61)$ distributions are sufficient to make an 
animated show of the distribution. We observed by this animation 
that the distribution changes systematically the shape with variation of either 
$\varepsilon$ or $N$ over the whole $N-\varepsilon$ grid 
though the onset of the apparent periodic distribution is quite 
rapid (a global check). 
We find that a special care must be taken over for the transient time when we are on the edge of 
the periodic window region. There the maps may fall sometimes into the final periodic attractor after,
for instance, $8 \times 10^4$ iterations but sometimes only after $10^2$ iterations depending on the
random initial values.  For this edge region we have done a separate analysis using 
a sequence of runs, each consists of $10^5$ to $10^6$ iterations. We comment on this region 
in Sec.  V  D.  
This simple but tedious analysis\footnote{ 
A GCML run can be done by a personal computer with a Pentium 300MHZ CPU 
in a day but for EGCML we used a hundred PCs(Pentium 90MHZ) simultaneously for a week
during Christmas in 1996. Then we assigned ranks of $1525$ distributions for both models.
}  was the start of the work  to unveil the real feature of the turbulent regime.

\subsection{The MSD surface}  %Fig.8a and 8b
In Fig.~\ref{fig8}a and b we show the MSD of the mean field ($h_n$) distributions at $a=1.9$ 
as a surface plot over the $\varepsilon-N$ grid for GCML and EGCML respectively. 
Both MSD and $N$ are plotted in the logarithmic scale so that the linear edge of 
the surface in the front panel at $\varepsilon=0$ is of course due to LLN.
For a non-zero but very small $\varepsilon$ $(\le 0.01)$,
the LLN still holds to a good approximation but for a  larger $\varepsilon$ we can clearly see that 
the surfaces have many peaks.
The global features of the MSD surfaces may be summarized as follows.

{\bf The valleys}~: At the valleys of the surface 
the MSD still approximately respects LLN and we may call these valleys as genuinely turbulent regions. 

Apart from the valleys LLN does not hold.
This does not imply the real violation of the LLN in the ensemble average\cite{pikovskyb}. 
What is meant by the violation of LLN here is that there is a larger fluctuation in the time series of $h_n$ 
than expected by LLN.  
We are interested in detecting the coherence among the evolving elements by the enhanced MSD. 
For both GCML and EGCML the peaks divide into two classes
except that the peaks are more parallel each other for GCML.
Below we quote figures for GCML.

{\bf The first-class peaks}~: We can see two very high and broad peaks at 
$\varepsilon \in 0.032 - 0.034$ and $\varepsilon \in 0.042 - 0.054$ and LLN is violated
extremely on them. Here the excess MSD is observed also for small $N$.
Let us call them as the first-class peaks of the MSD surface.

{\bf The second-class peaks}~: Separated by the large first-class peaks there are also several minor peaks.
Their locations for $a=1.90$ are
\begin{eqnarray}
\varepsilon &\approx&  0.012, 0.016, 0.020, 0.024   \label{second}\\
\varepsilon &\approx&  0.062,0.07,0.074,0.08,0.088-0.098, 0.106- 0.108 \cdots \nonumber
\end{eqnarray}
Let us call them as the second-class peaks. 
Provided that one keeps eyes only on the second-class peaks
the MSD first decreases with increasing $N$ following LLN and then above certain threshold $N_\delta$ 
the MSD ceases to decrease and remains approximately in a plateau.
This is a characteristic behavior of MSD at the hidden coherence.
However we show below that the $h_n$ distribution is not a gaussian but is a distorted gaussian 
in the bulk of the second-class peak 
regions. 

\subsection{The mean field distributions} %Fig.~\ref{fig9}a and 9b

Now we analyze directly the table of $25 \times 61$ distributions of the mean field
$h_n$ at $a=1.90$ for both GCML and EGCML to understand how the distribution looks like for 
the enhanced MSD regions. 
Before this analysis we had thought from the literature\cite{kb} that the distributions are all gaussian 
with possible MSD enhancement at least for the central region of the $h_n$ histogram
\footnote{
Since the $h_n$ distribution is limited in the range $[-x_L,x_L]$, it cannot be a finite width gaussian 
distribution over $(-\infty,\infty)$.  When we discuss whether the $h_n$ distribution is gaussian or not we concern whether 
the essence of the CLT,  that the convolution of  independent random distributions peaks 
like gaussian in the central region,  is in action or not. 
The ratio of the fourth to the second moment must be three for a perfect gaussian
and for the $h_n$ distribution a small  deviation from three is reported \cite{kb}.  
But it is sensitive to the tail region and the dynamics in the central region is obscured in a ratio. 
For instance, the rank three distribution below is often two overlapping gaussian peaks 
due to the periodic clusters with high mixing but a ratio of moments cannot tell it.  
 }.
But to our first surprise we find that 
there are a variety of distributions involved in the table of GCML and EGCML.
 
We assign each distribution a pattern classification number or a rank 
from zero to four by its shape.  
\begin{enumerate}
\item[0: ]   The distribution is gaussian. Compared with the gaussian 
distribution at $\varepsilon=0$ with common $a$ and $N$, the MSD agrees
within 20 percent error.

\item[1: ]   Still gaussian but with more than 20 percent increase of the MSD.
At large $N$ we see the MSD can become even more than twice.

\item[2: ]   A singly peaked distribution but the shape is distorted from gaussian.
Typically a gaussian-like peak polarized to the right or left, or trapezoidal. 

\item[3: ]   Either it has a few sharp peaks on top of a broad band, 
or it is an apparent overlapping gaussian distribution. 
It manifestly shows the periodic motion of elements. 

\item[4: ]   The distribution consists of a few sharp peaks only.

\end{enumerate}
At rank zero the mean field distribution obeys both LLN and CLT 
and the maps may be thought as independent random numbers with a common probability
distribution. Oppositely at rank four the maps are in periodic motion and so is the mean field.
The ranks are organized in a way that the periodicity of the elements
becomes more manifest with an increase of the rank.  

In Fig.~\ref{fig9}a and b we show the rank distribution on the $N-\varepsilon$ grid as a density plot  
for GCML and EGCML respectively.  
We find the followings.

(1) There is a remarkable match between the MSD surface plots 
and the distribution density plots\footnote{ 
In order to make a \lq blind test\rq~  we never compared the ranks with 
the MSD surface before we finish the whole task of rank  assignment.}.  
The MSD surface is high (low) wherever the rank of the distribution is high (low).
The rank distribution plot is almost a contour plot of the MSD surface.
This is highly non-trivial. That the rank plot is a contour plot of a surface 
means that there occurs no jump of ranks when we vary either $\varepsilon$ or $N$.
For instance to go to rank four from rank zero one has to pass the intermediate rank regions. 
This indicates that the dynamics of maps changes continuously (not necessarily smoothly) 
when either  $\varepsilon$ or $N$  is varied over the turbulent regime.

(2) The fist class peaks of the MSD surface occur just at the rank three and 
four distributions. The first-class peaks are due to the periodic coherent motions of 
elements. 

(3) In the second-class peaks the mean field distribution is rank zero (both CLT and LLN hold.)
for small $N$. When $N$ is increased with $\varepsilon$ fixed,  it becomes rank one (CLT holds but LLN violated ) 
above certain threshold.   
With a further increase of $N$ it almost always become rank two (even CLT is not fully operative). 
No higher rank distribution (rank three or four) appears in the second-class peak regions. 
Only in small band(s) of $\varepsilon$  (for instance, only in a narrow band around $\varepsilon=0.07$ for $a=1.90$)
the rank remains one for large $N$. 

(4) The  threshold between rank zero and rank one distribution precisely agrees with 
the break point $N_\delta$ of the drop of MSD following LLN.
For reference we superpose the approximate curve for the 
$N_\delta$ ($N_\delta \propto 1/\varepsilon^2$) given by \cite{kb} 
in Fig.~\ref{fig9} and indeed it runs roughly between the rank zero and rank one regions.
We consider the fact in (3) that the rank one distribution changes smoothly to the rank two distribution 
with increasing $N$ is a key to the hidden coherence; it indicates that the coherence between maps 
\lq hidden\rq~  in the rank one distribution region becomes manifest in the adjacent rank two distribution region.  

Hereafter we focus our attention to GCML. In Fig.~\ref{fig10} we show a sample of  (3) in the left three boxes.  
We fix $a$ and $\varepsilon$  ($a=1.90, \varepsilon=0.08)$ and give the $h_n$ distributions 
at various $N$ ($N=30,~100,~3000$) which are sampled in $10^5$ iterations from random 
starts discarding the first $10^4$ steps.  
We also display in each box the $h_n$ distribution with $\varepsilon$ set to zero as a reference distribution.  
In the box for $N=30$ the two distributions are indistinguishable.
The $h_n$ distribution is hence rank zero and both LLN and CLT hold. At $N=100$ the distribution
with $\varepsilon=0.08$ becomes a gaussian with enhanced MSD. It is rank one and indicates
the \lq hidden\rq~coherence; CLT holds but LLN does not\footnote{ A figure corresponding  to 
these two boxes was given in Fig.1 of  ref.\cite{kb}  (At $a=1.99$ and $\varepsilon=0.1$). 
But the continuous change of dynamics to higher ranks, such as from the second box to the third box was not noted.}.  
Now in the third box at $N=3000$ the distribution turns out to be the rank two distorted gaussian distribution.
Even CLT no longer holds.  In the fourth box we also exhibit the $h_n$ distribution at $a=1.99$, $\varepsilon=0.08$
and $N=1000$. This is in our classification the rank three distribution with two overlapping gaussian peaks.
In the model with $a=1.90$,  we can move to this dynamics region by varying $\varepsilon$, i.e. by  traversing 
the turbulent regime. This point is further substantiated in the following subsections. 

At this point we need some explanation on the terminology of the {\it hidden} coherence. 
The interesting phenomenon that above some $N_\delta$ the LLN in the time series is violated
certainly occurs but now we have found that it is restricted to the second-class peak regions only.
There are valleys where LLN approximately holds even for large $N$ and also the first-class peak regions
where LLN is violated even for small $N$. And now even in the second-class peak regions the place where
the CLT holds (with LLN violated) is again restricted. For $a=1.90$ it is only a narrow band around $\varepsilon=0.07$
and  it moves to the larger $\varepsilon$ for larger $a$. 
Thus if a distinction between the rank one and two distribution is made and the hidden coherence is defined 
strictly as the phenomenon that CLT holds with no approximation (i.e. excluding the rank two distribution)
but with broken LLN,  it can live only in a very restricted region among the second-class peak regions.
We will further show below that a careful analysis of the decay exponent of the temporal correlator can also 
unveil the coherence among maps. 

But still we consider the first discovery\cite{ka} of the onset of the violation of LLN {\it even 
in the region where the CLT holds (rank one)}  is remarkable.
Our point in emphasizing the distinction between rank one and two distributions is that the recognition that 
the rank one dynamics region is adjacent in the $N-\varepsilon$ space to the rank two dynamics region 
which in turn is adjacent to the rank three and so on enables us to track down the change of dynamics and 
thus leads us to a working hypothesis that the hidden coherence is the most modest manifestation 
of the coherent quasi-periodic motion of maps ubiquitous in the turbulent regime. 

\subsection{Traversing the GCML turbulent regime} %Fig.11 
Now that we have identified the various regions in the turbulent regime with distinct manifestation of 
periodicity let us to proceed to  clarify  how the dynamics of maps in these regions are related each other. 
To collect necessary information it is best to traverse the  turbulence regime at fixed $a, N$ and varying $\varepsilon$
as a tuning parameter.  In Fig.~\ref{fig11} we present  a sample record of such an expedition at $a=1.90, N=1000$. 
The first column shows the mean field distributions as well as the single map distributions.
The former is the average during the steps from $n=10^4$ to $n=10^5$, and the latter is the average during the last $2 \times 10^3$ 
steps of $10^5$ iterations.  In the second column  the rank of the $h_n$ distribution is denoted at the top in each box. 
The evolution of a hundred selected  maps is displayed for the last seven steps. 
In order to chase the periodic motion of elements 
we devise a simple temporal correlator of  the {\it relative coordinate vector}
$\tilde{\mbox{\boldmath $x$}}(n)\equiv (x_1(n) -h_n, \cdots, x_N(n)-h_n)$ 
of maps around the mean field:
\begin{equation}
  C(t)=\Biggl\langle  
          \frac{\tilde{\mbox{\boldmath $x$}}(n+t)\cdot \tilde{\mbox{\boldmath $x$}}(n)}{
         | \tilde{\mbox{\boldmath $x$}}(n+t)| | \tilde{\mbox{\boldmath $x$}}(n)| }   \Biggr\rangle . \label{correlator}
\end{equation}
The average over $n$ is taken for  the last $10^3$ steps.
This correlator  is displayed in the third column.  We show it just as defined by  (\ref{correlator})
but for the measurement of the decay exponent we quote values obtained by the signed
logarithm plot ($f(x) \equiv (x/|x|)$log$(|x|)$) of $C(t)$.  For smaller lattice analysis  we also averaged over data 
with different initial configurations to improve the statistics but this did not cause any significant change for
the exponent estimation.
The last column shows the return maps of $h_n$ (the set of 
points ($h_n,h_{n+1})$)  from the last $10^3$ steps.

Let us start at pure randomness $\varepsilon=0$. 
The map average distribution has many sharp peaks with fractal hierarchy 
structure reflecting the unstable fixed points of a single map in the chaotic region.
The $h_n$  distribution obeys of course both LLN and CLT and naturally sharp gaussian (rank zero)
for  $N=1000$. The evolution plot shows a simple logistic pattern and the return map is a simple tiny cloud of 
points. The correlator is one at $t=0$ by definition and decays almost instantly. 

We can foresee that periodicity three window in the first-class peaks is waiting us ($\varepsilon \in [0.032,0.054])$
but let us choose  $\varepsilon=0.028$ to see if  any  precursor of it emerges. 
We find that  the $h_n$ distribution has become rank one.  The map average distribution has lost some sub-peaks
since any one of the  maps  is pulled by the mean field in fluctuation and it has become more difficult for
 it to stay at the unstable sub-fixed points for long time.
The coherence is still invisible neither in the map evolution nor in the return map. 
But we find that remarkably the correlator now oscillates in period three and decays 
exponentially ($\approx 30$ steps for $10^{-3}$ decay).
This period three damping oscillation of the correlator starts 
at $\varepsilon \approx 0.02$ except for the valleys and the decay exponent becomes smaller for larger $\varepsilon$. 

At $\varepsilon=0.030$  the $h_n$ distribution is  rank two. We are now very close to the first-class peaks.  
The correlator survives longer period ($\approx 70$  steps for $10^{-3}$ decay) in period three motion.   
But the return map does not sense this. From the evolution diagram we can immediately tell that this failure of the return map 
is due to mixing of the maps between the quasi-clusters at a high rate. This inadequacy of a return map should be noted.  

At $\varepsilon=0.032$ we are on the edge of the first-class peak regions. 
Now very careful analysis is necessary to account for the first transient behavior of the maps.  
The truncation of the first $10^4$  steps is not sufficient and we have to watch 
the evolution carefully until at least during $10^5$ steps.  
We show two typical samples at  $\varepsilon=0.032$.  

In the first sample at $\varepsilon=0.032$ the maps remain in a few quasi-clusters with mixing until the end of the run. 
The unstable quasi-clusters oscillate in period three and the correlator decays to $10^{-3}$ only after 
 $t \approx 140$ due to the smaller mixing than before.  Now we can see in the return map three clouds of points turn out. 
The map average distribution has only three prominent peaks and all other sub peaks are lost. 
The corresponding $h_n$ distribution is rank three with both a visible peak and a shoulder on top of a broad band.
The high MSD of the $h_n$ distribution comes from the large separations between the dominant unstable 
fixed points of the maps in interaction. This distribution is an example of overlapping gaussian distributions. 
For the perfect periodic motion of the tight clusters,  sharp delta peaks would be observed in both map and $h_n$
distributions. Due to the decay of the quasi-clusters caused by the mixing the sharp delta peaks become broad and 
turn into overlapping gaussian distributions.  We frequently observed more prominent overlapping 
two or three gaussian peaks in the whole analysis of the turbulent regime using both $a$ and  
$\varepsilon$ as tuning parameters.
(See, for instance,  the fourth box in Fig.~\ref{fig10}.) 
 In the second sample at $\varepsilon=0.032$ the maps literally dropped into period three 
 attractor of three clusters at $n \approx 8\times 10^4$. 
 As the map evolution plot shows the map positions in each cluster 
 exhibit some variance in evolution so that we would better describe the clusters as quasi-clusters.
The broad band in this rank three $h_n$ distribution is an artifact of the transient motion
before the maps fall into the final attractor. There is no mixing at all between the clusters.
The $h_n$ distribution is a sharp single peak  and  the $h_n$ return map is confined  in a very small range.
This is due to the fact that the three clusters happen to be almost equally populated in this run so that the mean 
field fluctuation is minimized.

 At $\varepsilon=0.042$ we are at  the summit of the first-class peak. 
 Here  the maps almost always drop into period three attractor of two clusters.  
 By only two clusters in period three motion the mean field must also fluctuates in period three. 
 Thus the $h_n$ distribution has sharp three peaks and MSD turns out high. 
 The value of MSD is solely determined by the dynamics of clusters and $N$ independent. 
 Also the return map is widely spread. 

 While we  traverse the summit of the first-class peaks till $\varepsilon \approx 0.52$ 
the attractor remains in period three.  Sometimes it bifurcates to period six.   

In further expedition we depart  from  the first-class peaks and go down into the second
class peak regions.   Everything then occurs in reverse way in this descent till $\varepsilon \approx 0.058$.

From $\varepsilon \approx 0.06 - 0.1$ we traverse the second-class peaks as well as the valleys. 
At the second-class peaks the correlator survives above $10^{-3}$ until $t\approx 30 - 40 $ 
but with no apparent low periodicity.  
Then above $\varepsilon \approx 0.10$ the correlator starts again catching the precursor 
of the period two opposite phase 
motion in the two-cluster regime of the GCML.  Thus the indeterminacy of the period of quasi-clusters 
beyond second peak regions comes  from the fact that the regions are just in between the period three 
window and two-cluster opposite-phase motion regions.
After $\varepsilon \approx 0.1$ the correlator shows perfect period two motion and the decay is smoothly prolonged 
with increasing $\varepsilon$. At  $\varepsilon \approx 0.15$ it only decreases by a few percent even at $t=500$.
Correspondingly, above $\varepsilon \approx 0.1$ the middle peak in the map average distribution starts
approaching to the peak at higher $x$ and above $\varepsilon \approx 0.13$ the map average distribution
splits into two (no events in the middle gap region). Still in the evolution plot we can see the maps are in loose clusters 
with mixing between each other. The high MSD of the rank two and three distributions above $\varepsilon \approx 0.1$ 
is clearly induced by the widely spread dominant unstable fixed points of low periodicity. 

\subsection{The dynamics in the turbulent regime} %Table 1
We have done a similar analysis at various $a$ up to $a=1.99$ 
and verified that all above features at $a=1.90$ are unchanged. 
The various manifestation of the periodicity is summarized in Table \ref{table1} for $a=1.90$.
Only a minor difference for larger $a$ is that the regions of rank one distribution above the first 
rank MSD peaks become larger shifted to larger $\varepsilon$ and also the rank three distribution starts to appear there.
Since the rate of the mixing of maps between clusters is the key to the turbulent regime 
we have also measured the temporal correlator of the clustering pattern matrix similar to the 
Anderson-Edwards order parameter\cite{kb}. 
It exhibits the same variation of the decay exponents with our correlator.  
We examine below how the dynamics in various regions are related each other.
%------------------------------------------------------------------------------------------------

\noindent
(1) Let us first compare the dynamics of the periodic regions with that at $\varepsilon \approx 0$.
At $\varepsilon \approx 0$ maps evolve almost independently with long visit at all of the unstable 
fixed points of the logistic map. Thus the map average distribution has many peaks with 
fractal structure peaking. Since the maps behave as almost independent random variables 
with the same probability distribution, naturally both LLN and CLT hold.
At the periodic window in the first-class MSD peak regions, on the other hand, 
the $1-\varepsilon$ focusing effect tunes the maps to form tightly bound two or three clusters 
mutually oscillating in period three attractor and this leads to the rank four distribution
with sharp delta peaks. The dynamics of clusters in the periodic attractor is solely determined by 
$a$, $\varepsilon$ and the effective couplings between the clusters determined by 
the population {\it ratios} among the clusters\footnote{
We have numerically checked that the matrix-coupled three-map model similar to (\ref{matrix}) does exhibit 
the same period three attractor at the same value of $\varepsilon$.}. 
The MSD of the $h_n$ distribution is again determined by the $a$, $\varepsilon$ and the 
population {\it ratios} between the clusters, and there is no room for the explicit $N$ dependence.
Similar coherent dynamics is realized also in the ordered two-cluster regime beyond the turbulent 
regime ($\varepsilon > 0.2$ for $a=1.90$). 
The turbulent regime has a period three window in the center of it and the two-cluster regime beyond it.
We discuss below how these affect other regions.  

\noindent
(2) When $\varepsilon$ deviates from the range necessary for the period three cluster dynamics (or for the period 
two-cluster dynamics), the perfect synchronization of maps among each cluster is no longer realized.
But what we observed in the last subsection is that even for some large deviation of $\varepsilon$ 
the maps still move under certain coherence if $N$ is beyond the threshold $N_\delta$.
Such coherent motion of maps can be best described by {\it unstable quasi-clusters}
the decay life of which is given by the inverse of the exponent of the correlator.
The maps now divide into a few masses {\it at any instance} of the evolution. 
In a short time scale these masses oscillate mutually in the same periodicity as before 
as we observed in both the map evolution plot and the correlator\footnote{
Note that the quasi-clusters are hardly seen in the return map analysis because of the mixing.}.
But the variance of map positions among each mass is not so small and the synchronization of maps 
in each mass is also not perfect. Hence we call these masses as quasi-clusters. 
Due to the fail of perfect synchronization, the quasi-clusters mix each other by the exchange of maps 
and soon they loose their identities. So the quasi-clusters are unstable and have short life time. 
But still at any instance of evolution there do exist quasi-clusters. 
The quasi-clusters defined at a certain time decay but in the evolution of maps new quasi-clusters 
are kept created. We may regard these quasi-clusters as {\it remnants} of the perfect periodic motion 
of maps.
In fact when we approach the periodic regions the decay exponent of the correlator decreases smoothly, that is, the 
life time of the quasi-clusters increases smoothly, signaling the phase transition from turbulence to periodicity.

In pure turbulence the system is invariant under the exchange of maps,
while in the system of maps divided into clusters this permutation symmetry is  broken.
Simultaneously the system in pure turbulence is invariant under the temporal translation, 
while the system in periodic motion is only invariant under the temporal translation just for the period.
The self-organized state of maps in periodic quasi-clusters 
emerges as the state of broken symmetry from pure turbulence 
by a weak self-coupling and it emerges only for large $N$.
This is interestingly reminiscent of the onset of ordered parameter 
at the spontaneous break down of the global symmetry in the field theory 
where an infinite number of dynamical degree of freedom is necessary in order to 
support the ordered vacuum.  See, for instance, Weinberg\cite{weinberg} 
where the broken symmetry state of a chair is elucidated. 

It is important that the maps in the quasi-clusters approximately follow 
the previous cluster attractor in the life time of the quasi-clusters. 
This means that the $h_n$ distribution is controlled by the scale of the previous attractor approximately 
and again determined by the $a$, $\varepsilon$ and the population {\it ratios} and not by $N$.  
Thus there emerges a $h_n$ distribution whose MSD does not decrease with $N$
and LLN is violated.  
Depending on the rate of mixing and the allowed population ratios at the $\varepsilon$, distributions of various 
ranks emerge. 

\noindent
(3) If the deviation of $\varepsilon$ from the necessary range for periodic cluster attractor is not large, 
the failure in the map synchronization is not also large. 
The quasi-clusters decay by low rate mixing of members so the correlator survives for long time
exhibiting clearly the periodicity of the previous periodic cluster attractor. 
But the would-be delta-peaks of both the average map distribution 
and the $h_n$ distribution at no mixing now change into overlapping gaussian distributions 
with visible separation between the peaks.
These changes of the correlator and the $h_n$ distribution at small deviation of $\varepsilon$ 
are precisely what we observe around the first-class MSD peaks and near the upper boundary 
of the turbulent regime adjacent to the ordered two-cluster phase. 

With further deviation of $\varepsilon$ the mixing rate becomes higher. The correlator decays  
more quickly and the $h_n$ distribution becomes from rank three to rank two - the distorted gaussian distribution. 
We observe this when we go down from the first-class peaks to the region of $\varepsilon \approx 0$
or when we approach the first-class peak regions from above.
In the former case the correlator keeps exhibiting the period three damped oscillation even 
if we go down into the rank one $h_n$ distribution region. In the latter case 
the periodicity of the correlator becomes indefinite below $\varepsilon \approx 0.1$ 
for $a=1.90$ since naturally we are coming down from two-cluster regime with period two 
to the period three window, and we finally loose the track of a periodicity remnant (rank two distribution) when 
we go further into the narrow band of rank one $h_n$ distribution region. 
This is the genuine hidden coherence region.

\noindent
(4) The place the hidden coherence can live has now become very limited since we have improved 
the sensitivity to the coherence effect. 
In the old analysis\cite{kb}, the distinction between rank one and two distributions was ignored 
and neither the change of the decay exponent of the correlator with $\varepsilon$ nor 
the periodicity of it was analyzed. The periodic window and the valleys were not discussed also.
Then the coherence seemed to live hidden everywhere in the turbulent regime if $N > N_{th}$. 
Now, both the periodic window with manifest coherence and the valleys with valid LLN should be excluded
first of all.  Furthermore, in the regions where the correlator has clearly caught the periodicity remnant,  
it is no longer legitimate to describe the coherence as hidden. 
Thus, regarding the correlator the coherence is hidden 
only in a limited region in between the period-three and period-two remnant regions  
among the second-class MSD peak regions. This is the region with $\varepsilon \approx 0.06 - 0.1$ for $a=1.90$.
See Table \ref{table1}  and Eq. (\ref{second}). 
Here the correlator does show longer decay time ($10^{-3}$ decay within $20-40$ steps)  
than that at $\varepsilon \approx 0$ but it decays with large fluctuation showing no low periodicity. 
This is natural for the region in between two different periodicity regions.

But not only the correlator but also the shape of the $h_n$ distribution is a vital quantity to keep track 
of the change the dynamics. Almost everywhere in the possible residence of the hidden coherence 
regarding the correlator, the $h_n$ distribution is rank two (rather than gaussian) and naturally understood 
as the smooth continuation of the rank three overlapping gaussian distribution. 
The CLT is apparently violated here and the distribution is still controlled by the scale of the attractor.

Therefore, the place where the coherence is literally hidden from both the correlator and distribution 
analyses is now very limited. It is the narrow band at $\varepsilon \approx 0.07$ for $a=1.90$ and around 
$\varepsilon \approx 0.09-0.11$ for $a=1.99$ where the $h_n$ distribution is really gaussian 
with enhanced MSD (rank one) and the correlator does not show the low periodicity.
Here the coherence seen in the enhanced MSD can be only substantiated either by the finite mutual information 
or by the instability of the FP fixed point distribution.  
But in (3) above we successfully tracked down the smooth change of distribution (and the corresponding change 
of the decay exponent of the correlator) with the deviation of $\varepsilon$ from rank four down to rank two 
distribution regions which surround the literally hidden coherence region.  
This gives a new way to look at the hidden coherence. 
Higher rate mixing with further smooth deviation of $\varepsilon$ should necessarily 
weaken the coherence among the maps and change the distorted gaussian (rank two) to the gaussian
distribution.  If the scale of the attractor still remains to govern the map dynamics the $h_n$ distribution 
with $N$-independent MSD would emerges. We propose this last statement as a working hypothesis 
for the origin of the hidden coherence.

\subsection{An Analytic Approach}

Up to this point we have heavily depended on the numerical simulation.  Our concern is the statistical properties of the large-size GCML (EGCML) and the Monte-Carlo simulation is naturally the basic approach to the phenomenology of such system. 
We have shown that a faithful extensive analysis of the data unveils amazing new features of the system.
But, of course, if by some analytical method, one could relate the systematic manifestation 
of periodicity in the turbulent regime to the periodic behavior of the element of the system, namely the 
period three window of a single logistic map, it would be certainly a step forward to the real understanding 
of the turbulent regime. In this last subsection we discuss such an analytic approach. 

Let us consider a typical case where the N maps are divided into equally populated three clusters $A,B,C$
($N_A=N_B=N_C=N/3$) and the clusters move cyclically round the three fixed positions $x_1,x_2, x_3$. 
In such a maximally symmetric and period three motion of GCML maps, $h(n)$, the meanfield of all maps
at time $n$ becomes a time-independent constant which is nothing but an equal weight average 
$h^*$ of $x_i (i=1,2,3)$.
To see this let us first rewrite the meanfield using the invariance rule (\ref{sumrule}) as
\begin{eqnarray}
   h(n) \equiv \frac{1}{N} \sum_{i=1}^N f(x_i(n))= \frac{1}{N} \sum_{i=1}^N x_i(n+1).
\end{eqnarray}
Then for the configuration in case the last average can be calculated as 
\begin{eqnarray}
  \frac{1}{N}  \sum_{i=1}^N x_i(n+1) = \frac{1}{N} \sum_{I=A,B,C} x_I(n+1)
= \frac{1}{3} \sum_{i=1}^3 x_i \equiv h^* 
\end{eqnarray}
where $x_I(n)$ denotes the coordinate of the cluster $I$ at time n.
Therefore, if such a configuration is produced by the interaction between the maps via the meanfield,
any one of GCML maps must be evolving by a common evolution equation
\begin{eqnarray}
    x_i(n+1)=(1-\varepsilon)f_a(x_i(n))+\varepsilon h^* ~~~(i=1, \cdots, N)\label{consthevolution}
\end{eqnarray} 
We can cast the evolution equation (\ref{consthevolution}) 
with a constant meanfield $h^*$ and $f_a(x) = 1 -a x^2$ to a simple logistic map 
\begin{eqnarray}
    y_i(n+1)=1- b \left(y_i(n)\right)^2  ~~~(i=1, \cdots, N ).
\label{yevolution}
\end{eqnarray} 
with a reduced coupling constant $b$  by a linear scale transformation \cite{PerezCerdeira}  
\begin{eqnarray}
    y_i(n) = \frac{x_i(n)}{1-\varepsilon+\varepsilon h^*}  \label{lineartransformation}.
\end{eqnarray}
The reduction rate $r$ is given by
\begin{eqnarray}
   r \equiv \frac{b}{a} = (1- \varepsilon)\left( 1 - \varepsilon (1- h^*) \right)  \label{rconstarint}.
\end{eqnarray}

The period three window of the logistic map (\ref{yevolution}) starts at $b=7/4=1.75$
by the tangent bifurcation and after the sequential bifurcation it closes at $b \approx 1.7903$ by the crisis.
%1.79035  
Therefore the reduction rate $r$ must be for $a=1.90$ in the range 
\begin{eqnarray}
               0.921 \le  r \le 0.942
\end{eqnarray} 
and at a given rate $r$ within this range,  (\ref{rconstarint}) gives a curve of constraint on the $\varepsilon - h^*$ 
plane.

There is another constraint from self-consistency.  Each of the maps $y_i$ is related to the GCML maps 
$x_i $ by the linear transformation (\ref{lineartransformation}) and so is the average value $y^*$ of maps  
$y_i$ to $h^*$:
\begin{eqnarray}
    y^* \equiv \frac{1}{3} \sum_{i=1}^3 y_i = \frac{h^*}{1-\varepsilon+\varepsilon h^*}. \label{yconstaraint}
\end{eqnarray} 
Since $y^*$ is simply an equal weight average of the period three stable orbits of ($\ref{yevolution}$),  
it can be estimated solely by the property of the logistic map without any recourse to the GCML evolution equation. For instance at the tangent bifurcation point ($b=7/4, r=0.921$), we have
\begin{eqnarray}
  f_{b}(f_{b}(f_{b}(y))) -y = b^6 (f_{b}(y) -y)\left(\Pi_{i=1}^3(y-y_i)\right)^2
\end{eqnarray}
and from matching of the coefficients we obtain $y^*=1/(3 \cdot 2 b) = 2/21$.
Numerically we find that $y^* \approx 0.25$ slightly below the threshold $(r=0.91)$, 
drops sharply ($y^* \propto \sqrt{b_{th}-b}$)  to $y^*=2/21=0.095$ at the threshold $(r=0.921)$, 
and $y^*$ gradually changes around $0.08$ until the end of the window ($r=0.942$).   
By eliminating $h^*$ from (\ref{rconstarint}) and (\ref{yconstaraint}) we obtain
\begin{eqnarray}
    \varepsilon = 1 - \frac{r y^*}{2}  - \left( r (1- y^* ) +  (\frac{ry^*}{2})^2   \right) ^\frac{1}{2}. \label{epsilonestimates}
\end{eqnarray}  
The estimated $\varepsilon$ for $a=1.90$ from (\ref{epsilonestimates}) is 
\begin{eqnarray}
   A: ~~   \varepsilon &=& 0.0523   \mbox{~~~at~~~}          (b, r, y^*)=(1.730, ~0.910,   ~0.250)    \nonumber \\
   B: ~~   \varepsilon &=& 0.0422   \mbox{~~~at~~~}          (b, r, y^*)=(1.750, ~0.921,   ~0.095)    \nonumber \\
   C: ~~   \varepsilon &=& 0.0365   \mbox{~~~at~~~}          (b, r, y^*)=(1.769, ~0.931,   ~0.070)    \label{estimates}  \\
   D: ~~  \varepsilon  &=& 0.0305   \mbox{~~~at~~~}          (b, r, y^*)=(1.790, ~0.942,   ~0.080)    \nonumber  
\end{eqnarray}
% A: e=0.0523201, r=0.91,                     (b=1.729)  
% B: e=0.0422164, r=0.921,    y*=0.095  (b=1.75)        period3 window starts at b=1.75,    (r=0.92105)
% C: e=0.0364547, r=0.93079, y*=0.07   (b=1.76851)  period6 starts at b=1.76851            (r=0.93079)
% D: e=0.0304678, r=0.94229, y*=0.08   (b=1.79035)  period3 window ends at b=1.79035 (r=0.94229)
The estimates A, B, C and D are evaluated respectively below the threshold, at the threshold, at the first bifurcation 
point at the middle, and at the upper edge of the period three window in the order of increasing $b$.

Before comparing with the simulation result we should note that a maximally symmetric configuration 
of equally populated three clusters in period three motion is assumed in deriving (\ref{epsilonestimates}).
Thus, the value of $\varepsilon$ necessary for such a configuration should be deduced from 
the genuinely period three region, namely from the interval from B to C. We can predict that 
$\varepsilon$ must be in the range (0.0365, 0.0422) for the formation of such a maximally symmetric GCML 
configuration. For such a configuration the meanfield should not fluctuate in time and it will show up 
as a deep valley in the GCML MSD surface.   We have added two exterior estimates in (\ref{estimates}) 
expecting that these extensions will cover the asymmetrically populated clusters in period three motion.

Now let us compare (\ref{estimates})  with the range of the period three first-class peaks found  in Section V B,
namely,  $\varepsilon \in 0.032 -0.034$ and $\varepsilon \in 0.042 - 0.054$.

Firstly there is a remarkable agreement between the range of peaks and the range of estimates A to D
 in (\ref{estimates}).  The whole period three window of a single logistic map shows up as the whole first-class
peaks.  A closer comparison with the observations in section V D gives further support for the link 
of the structure observed in the numerical simulation and the period three window of the logistic map.

\begin{itemize}
\item[1]
We noted that in the deep valley of the MSD surface between the two 
first-class peaks the formation of three clusters in period three motion.
The valley region $(0.034, 0.042)$ remarkably matches the above estimated range (0.0365,0.0422)
for the maximally symmetric configuration.

\item[2]
We noted that in the upper first-class peak ($\varepsilon \in 0.042 - 0.054$)  the maps almost always 
form only two clusters in period three motion. It is natural that the larger 
$\varepsilon$ reduces the number of clusters from three to two. As we noted the extremely high MSD 
in the first-class peak is induced by the asymmetric population configuration 
($N_A \approx N_B \approx N/2, N_C=0$).
\end{itemize}

Furthermore we have verified by simulation that the first-class peak regions shift to higher $\varepsilon$ 
when $a$ is increased precisely in accord with the prediction by (\ref{epsilonestimates}).
 
It is remarkable that the maps with high nonlinearity are driven into coherent synchronizing clusters with very small interaction parameter $\varepsilon$ once it takes the tune up value.
A lot of further effort has to be devoted to fully appreciate the implication.
The analytic approach here is only a step forward to this direction.
A survey of the overall relation of peak-valley structure of the MSD surface to the hierarchical windows of a single logistic map as well as the dynamical variation of the population configuration is now underway.

\section{Conclusion}

In this article we have revisited the GCML and first resolved the puzzle of the posi-nega switch
realized the two-cluster phase. 
The puzzle was that the maps posses the memory of the cluster to which they subject 
even though they pass through a transient  chaotic process where many channels are open. 
We have proved that maps never mix across the mean field of them in the transient process.
We have then analyzed the turbulent regime of GCML.  To date the so-called hidden coherence has been the main 
target of interest in this region. We have presented our new finding that there exists a remarkable 
period three attractor regions in this regime.  The turbulent regime locates in-between the genuinely 
turbulent region around $\varepsilon=0$ and two-cluster regime of period two attractor and 
has this period three attractor regions in the center of it.
We have shown that the remnant of the periodic coherence can be found almost everywhere in the turbulent regime
by the measurement of the temporal correlator of the fluctuation of  maps around the mean field 
and by the extensive analysis of the $h_n$ distributions over the whole turbulent regime.
The dynamics of the turbulent regime can be well described by the quasi-clusters
with mixing of elements between each other.  The hidden coherence may be regarded the most 
modest manifestation of this remnant periodicity. By an analytic approach we 
have shown that the prominent periodicity manifestation 
in the turbulent regime is caused by the period three window of a single logistic map.

A new extended globally coupled map lattice(EGCML)  is constructed in which the interaction between maps
decreases proportionally to the distance.  EGCML at $d=2$ has the same phase diagram 
with that of GCML.
The bifurcation tree structure of the attractor of EGCML in the two-cluster regime is also the same with that of GCML.
The maps in each EGCML cluster are phase synchronizing rather than synchronizing.
The posi-nega switch between the two clusters is realized also in EGCML.  
In the chaotic transient process in EGCML-switch,  formation of amazing spatial clusters of maps is observed.
EGCML may be viable for the simulation of chaotic diffusion process.
We have also shown that the turbulence regime of EGCML is also not immune from periodicity just like GCML 
turbulent regime.
As a whole this work is an exploration of order in the chaos. The posi-nega switch in GCML is realized 
because there occurs no mixing in the chaotic transient process. In the chaotic transient process 
in a EGCML-switch,  interesting spatial clusters are formed. And the turbulent regime of 
both of GCML and EGCML is dominated by the quasi-clusters of maps in periodic motion.

\acknowledgments
It is a pleasure to note the author's debt of gratitude to K. Kaneko.
This work was started several years ago stimulated by the pioneering articles on GCML 
by him\cite{ka,kb} and the compact model GCML has been an unfailing source of interest since then.   
The author is thankful to Fumio Masuda who worked on this subject together with the author 
until the last one year of completion and  the nice collaboration time with him is memorized.
Thanks also go to Hidehiko Shimada for many crucial suggestions, Hayato Fujigaki for his collaboration 
on the study of coupled flows done in parallel with this work,
and Kengo Kikuchi for his unfailing effort on the analytic approach to GCML now underway.
Last but not at least the author wishes to thank Wolfgang Ochs at the Max-Planck Institute 
for sharing his insight in this research and reading the manuscript.   

This work was supported  by the Faculty Collaborative Research Grant from Meiji University,
Grant-in-Aids for Scientific Research from Ministry of Education, Science and Culture of Japan,
and Grant for High Techniques Research from both organizations.

%%-----------------------------------------------------------------------------------------------------%%

\begin{figure}
\caption{
The transition from periodic attractors into the chaotic intermittent process. 
The GCML with $N=50$ and $a=1.90, \varepsilon=0.3$.  The threshold is $N_{th}=20$. 
The two clusters oscillate oppositely in phase until time $n_1$.  
Both clusters start the grand chaotic motion at $n_2$, i.e.  just after the minority cluster (dashed line) 
crosses the $x^*$ line. 
\label{fig4}}
\end{figure}

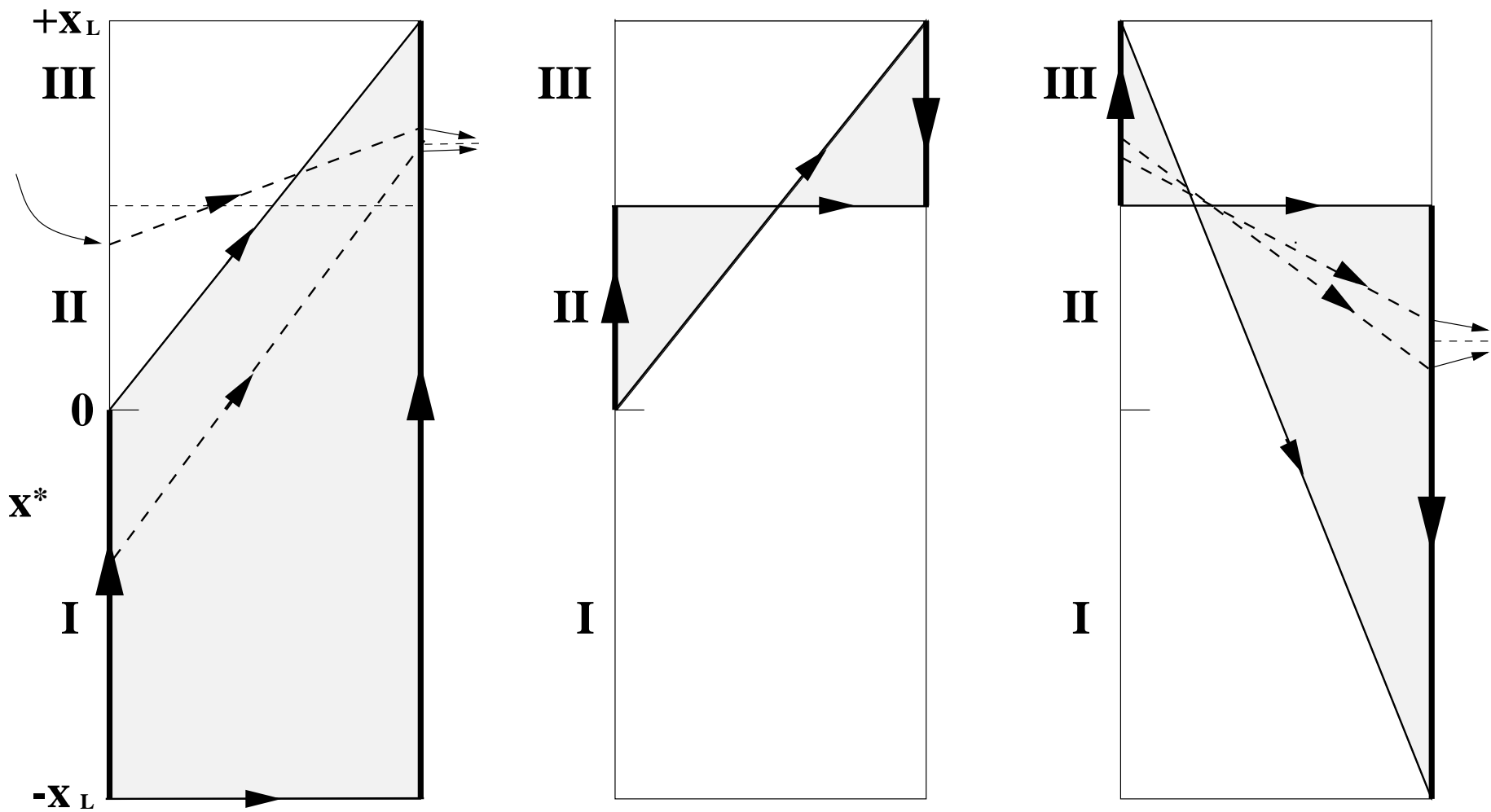
\begin{figure}
\caption{
The logistic map and transition from periodic to chaotic evolution. 
When a minority cluster is pulled down across $x^*$ from region III to II by the majority cluster, 
both clusters are mapped together into region III (the left box).  
Then after a small contraction by interaction they are again mapped together in the same direction 
and the quasi-coherent grand chaotic motion starts (the right box). 
\label{fig5}}
\end{figure}

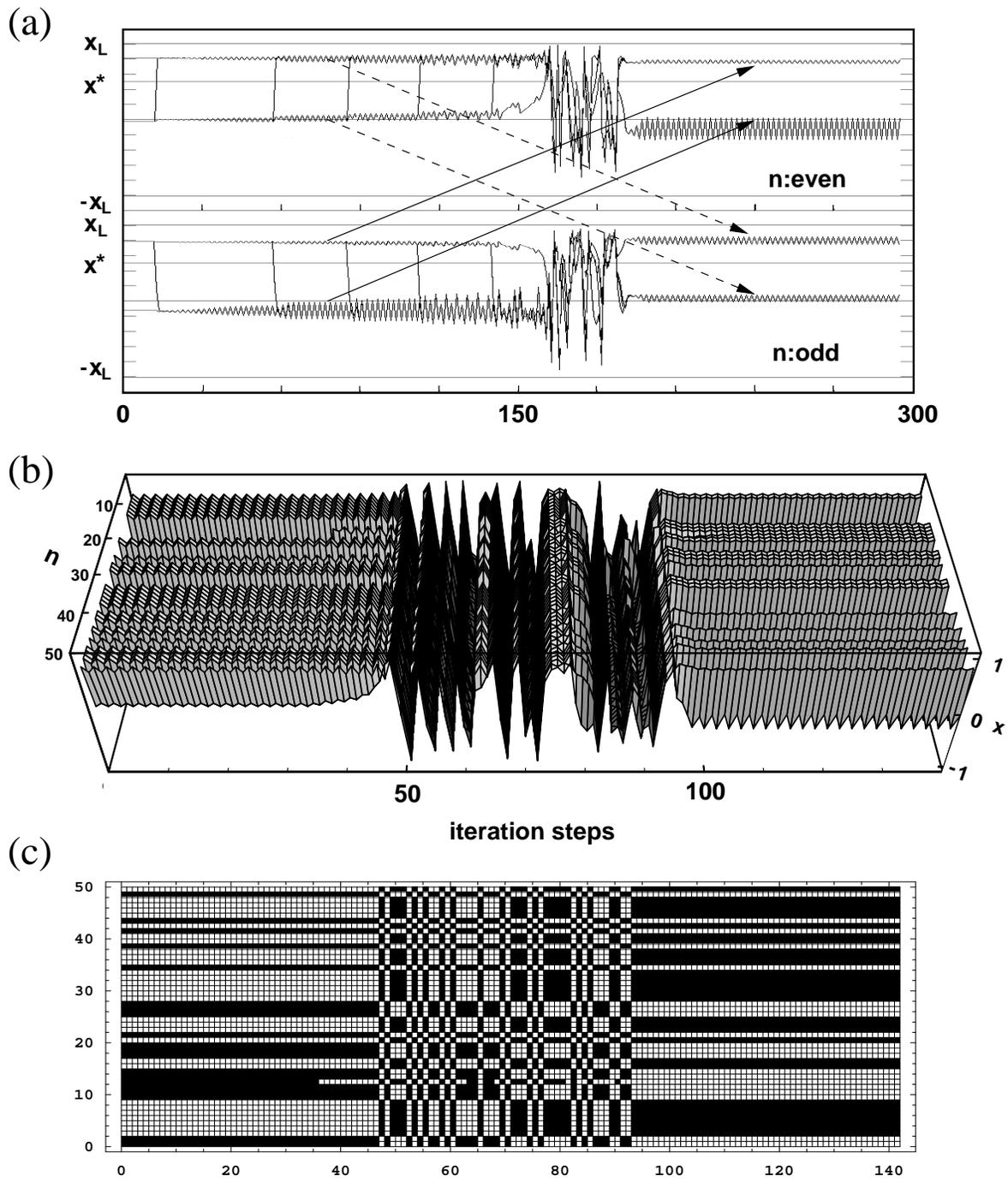
\begin{figure}
\caption{
The GCML posi-nega switch mechanism.   
The iteration steps are denoted in unit of two in all figures.
(a) The typical posi-nega switch in GCML ($N=60,~ a=1.98, ~\varepsilon =0.3$).
The attractor at even (odd) steps is shown in the upper (lower) diagram.
The attractor on the whole, the even and odd iteration steps together,  is precisely 
the same before and after the chaotic transition. 
The arrows point out that the posi-nega switch is a change in the count of the parity (even and odd) of iteration steps. 
(b) A typical posi-nega switch in GCML ($N=50,~ a=1.98, ~\varepsilon =0.3$). 
Evolution of maps at even iteration steps is displayed as a surface plot.  
The posi-nega switch is seen as the total swapping between 
peaks and valleys of the surface via the chaotic transient process. 
The chaotically oscillating surface is horizontal at any instance reflecting 
the quasi-clustering in the transient process.
(c) The same GCML switch with (b). The maps are distinguished by their mean field instead 
 of  $x^*$.  The print-circuit pattern shows no mixing of maps across the mean field.
\label{fig6}}
\end{figure}

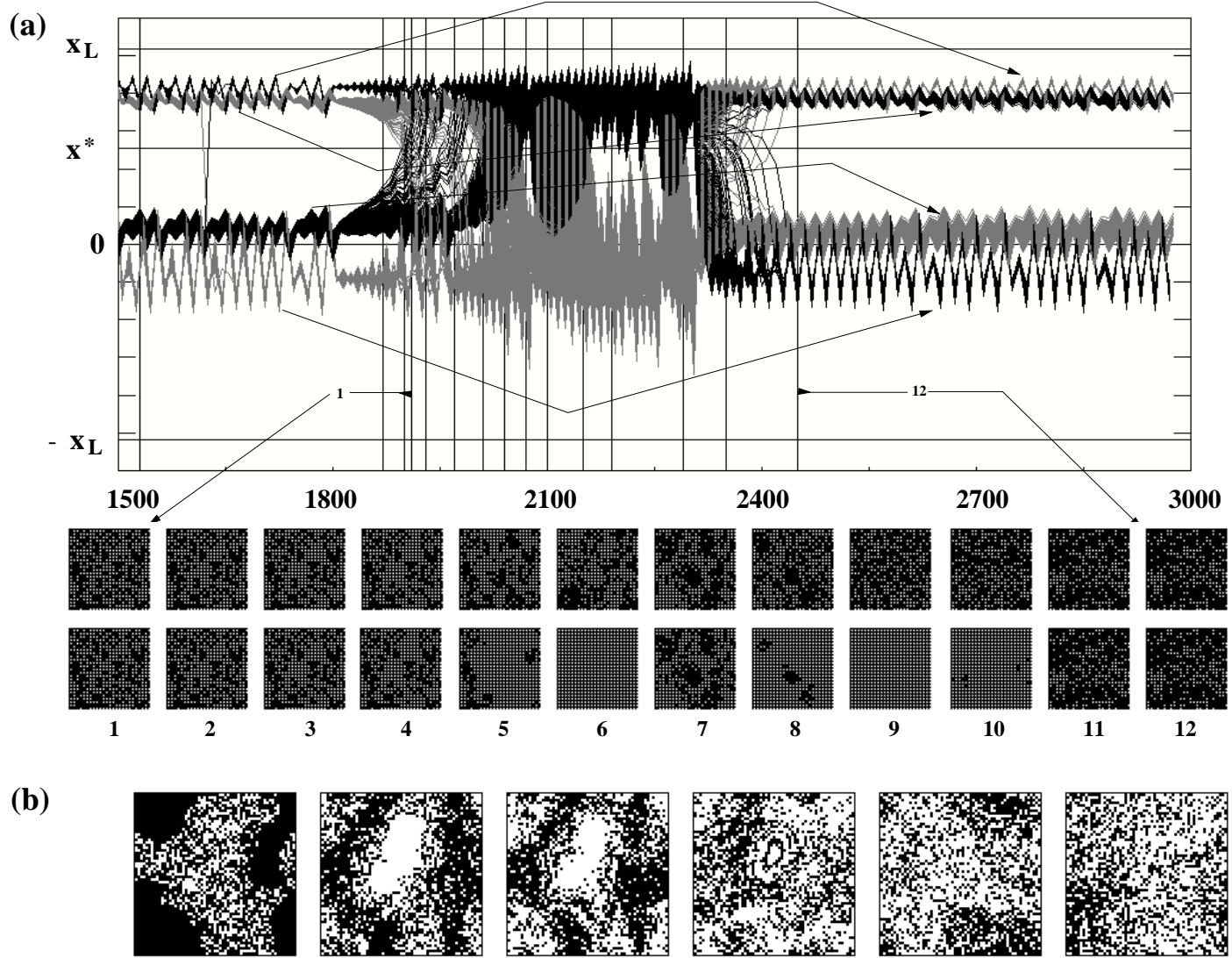
\begin{figure}
\caption{
(a) An EGCML switch with $25 \times 25 $maps ($a=1.9, \varepsilon=0.3$) 
from a random start. 
In the upper diagram, black (gray) lines connect maps at even (odd) steps and 
arrows indicate the switch. 
Lower boxes show the snapshots of the lattice at the even iteration steps 
indicated by the vertical bars respectively. 
In the upper array positive and negative maps are distinguished by the mean field 
$\langle\langle x(n) \rangle\rangle$  
while in the lower array by the $x^*$.  
The former exhibits the percolation formation of self-organized spatial clusters by synchronization. 
Between the snap-shots 4 and 8 the spatial clusters undergo posi-nega pattern switch.
(b) A sample of cluster formation in the large EGCML ($N=61 \times 61, ~ a=1.9, ~\varepsilon=0.3$) 
by the distinction by the mean field. 
A remarkable pattern like globular nebula is seen. 
\label{fig7}}
\end{figure}

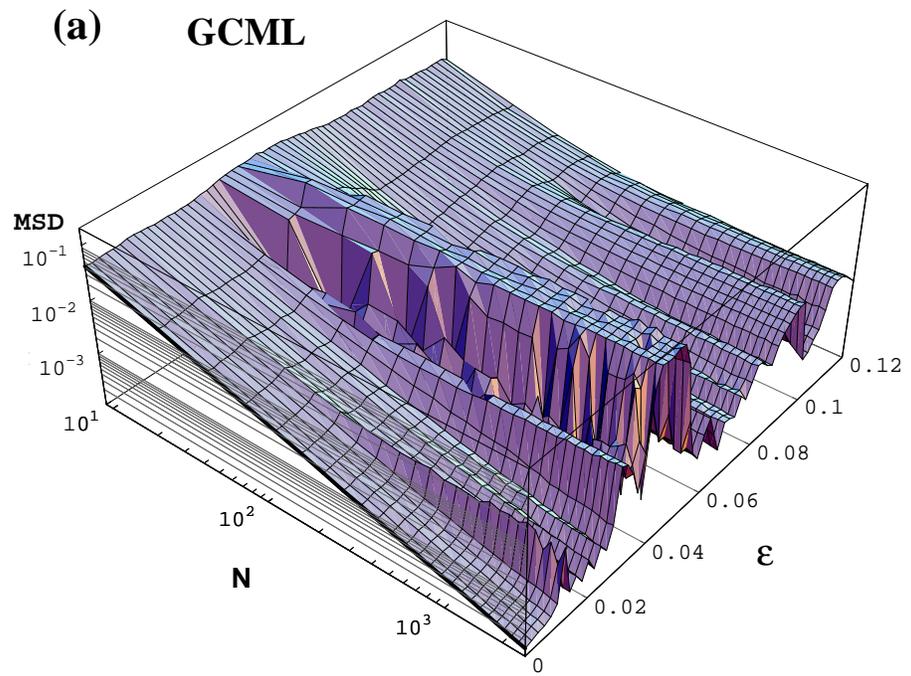
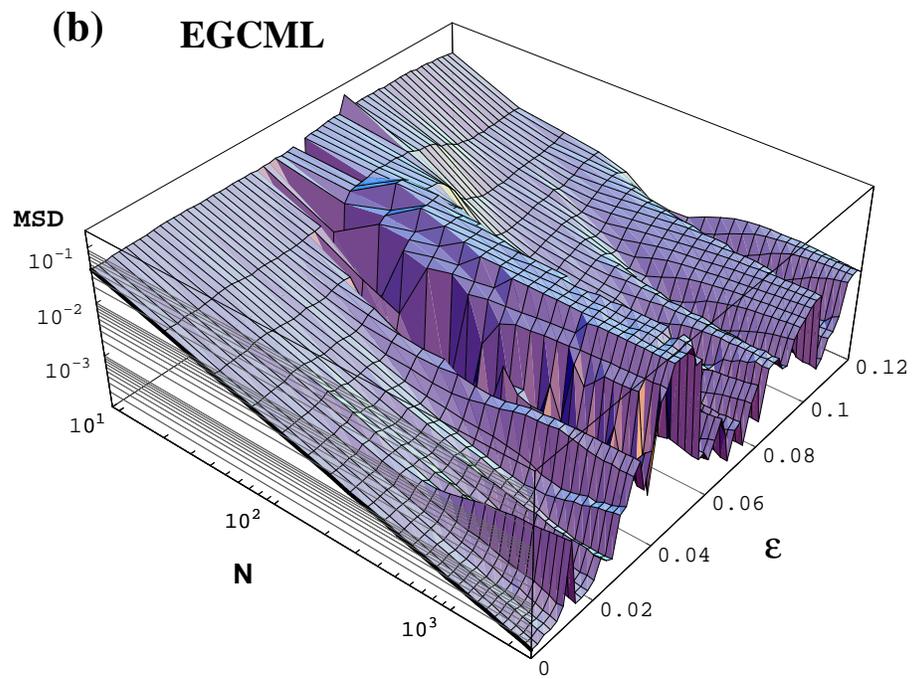
\begin{figure}
\caption{
MSD surfaces of mean field fluctuation  of (a) GCML and (b) EGCML at $a=1.90$ over the $\varepsilon-N$ grid. 
The peaks reflect non-trivial coherence between maps in the turbulent regime.
\label{fig8}}
\end{figure}

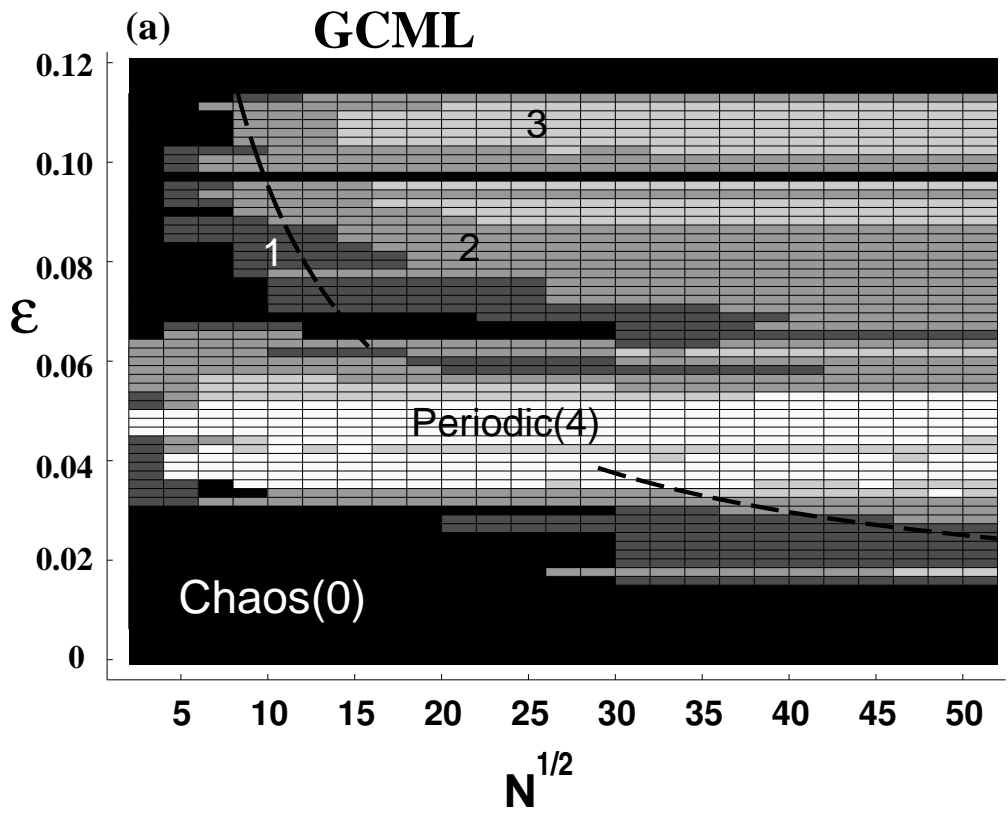
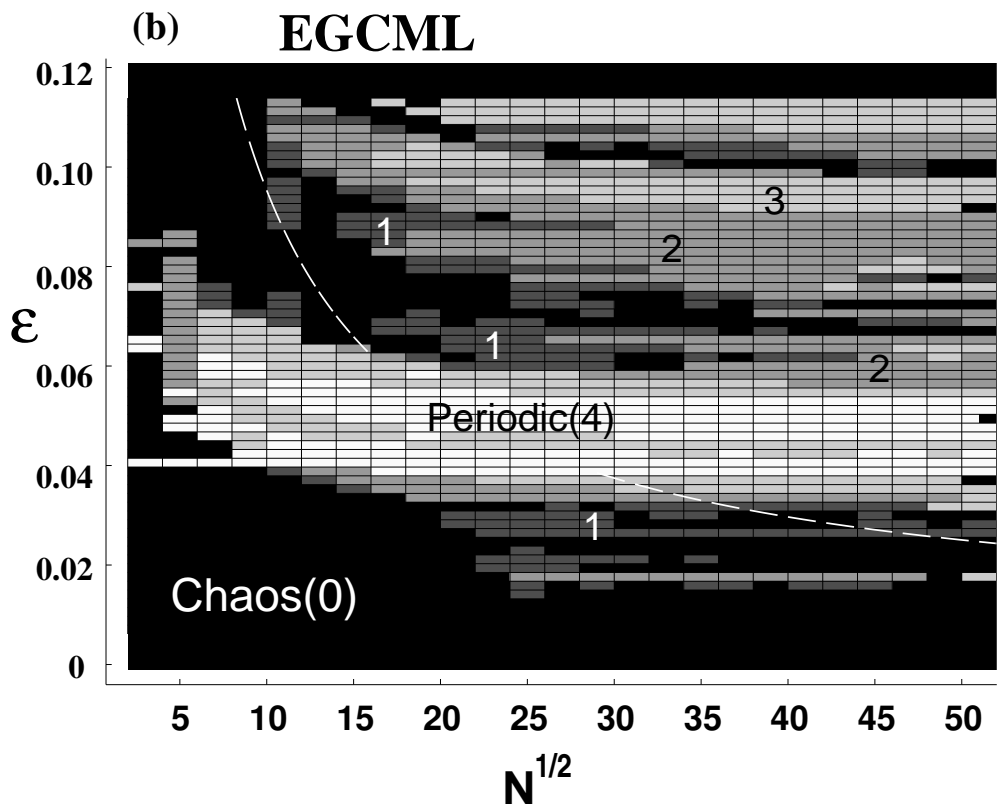
\begin{figure}
\caption{
The rank distribution on the $N-\varepsilon$ grid at $a=1.90$ as a gray-scale density plot.  
(a) GCML, (b) EGCML.  The rank varies from  zero(black) to rank four(white).  
An approximate estimation curve for the threshold $N_\delta$ ($N_\delta \propto 1/\varepsilon^2$) 
is superposed. 
\label{fig9}}
\end{figure}

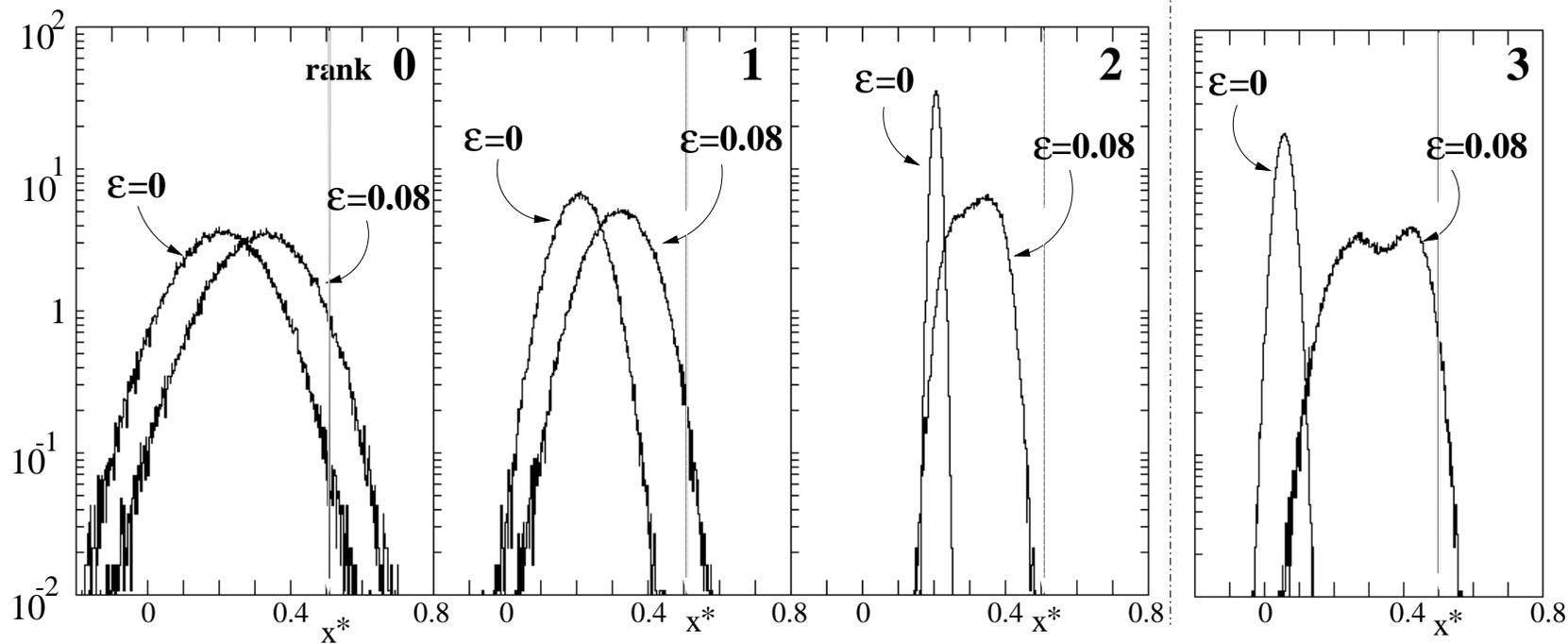
\begin{figure}
\caption{
The GCML $h_n$ distribution sampled in $10^5$ iterations from random 
start discarding the first $10^4$ steps using 1000 bins.  
In each box the $h_n$ distribution with the rank indicated at the corner is compared with the reference 
distribution at $\varepsilon=0$.
The left three boxes:  $N=30,~100,~3000$, $a=1.90$ and $\varepsilon=0.08$. 
The fourth box:  $N=1000$, $a=1.99$, $\varepsilon=0.08$.     
\label{fig10}}
\end{figure}

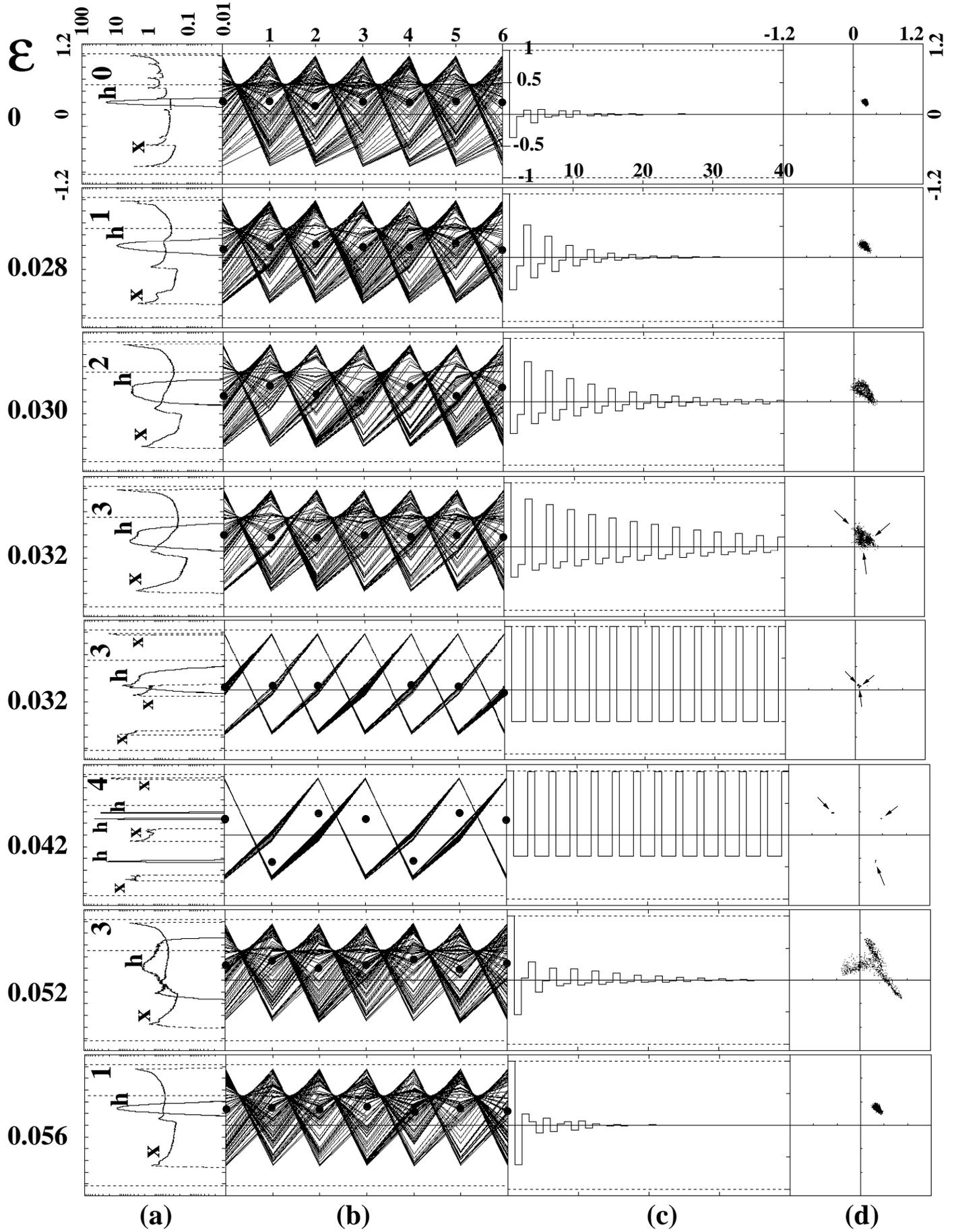
\begin{figure}
\caption{$N=10^3$ GCML with $a=1.90$.
The variation of dynamics with the change of the coupling $\varepsilon$ through the
first-class MSD peak regions. All quantities in a row are measured in the same run ($10^5$ iterations)
and the scales for each column are given in the top box at $\varepsilon=0$.
For $\varepsilon=0.032$ two sample runs are shown; in the lower, maps fell into a three cluster attractor in 
period-three motion after $8\times10^4$ iterations; in the upper, they survived in quasi-random motion 
over $10^5$  iterations.
(a) The mean field distribution (marked as $h$) sampled discarding the first $10^4$ transient steps
and the map distribution ($x$) averaged over the last $2 \times 10^3$ steps. 
(b) Clustering pattern. The  lines show the evolution of randomly selected 100 maps for 
the last seven steps and the black circle is the mean field of all $N=1000$ maps. 
(c) Temporal correlator between the two relative-coordinate vectors of maps. 
Averaged over the last $10^3$ steps. Below and slightly above the first-class peak regions
(the periodic window) it decays exponentially exhibiting the period three motion of quasi-clusters. 
Approaching the periodic window the decay exponent smoothly decreases.
(d) The return maps for the same period as (c). Less sensitive to the quasi-clusters.
The arrows indicate the visible period three quasi-clusters.
\label{fig11}}
\end{figure}

\mediumtext
\begin{table}
\caption{The various periodicity manifestation in the turbulent regime of GCML ($a=1.90$ and $N=1000$).  Note that in each range of $\varepsilon$ there are also valleys with an approximate LLN 
along with listed peaks with the broken LLN.
\label{table1}}
\begin{tabular}{cccc}
$\varepsilon$& region name&rank of $h_n$ distribution\tablenotemark[1] & correlator periodicity \\
\tableline
$0.01  -   0.032$ & the second-class peaks & $ 1  ,2   $ & $[3]$ \tablenotemark[2]         \\      
$0.032 -  0.054$ & the first-class peaks       & $ 3  ,4   $ & $ 3 $ \tablenotemark[3]         \\     
$0.054  -   0.06$  &$\updownarrow$\tablenotemark[5] & $ 1  , 2  $ & $[3]$ \tablenotemark[2]         \\
$0.06   -    0.1$   & the second-class peaks & $ 1  , 2  $ & $[?] $ \tablenotemark[4]         \\
$0.1     -  0.15$   & the second-class peaks & $ 1  , 2  $ & $[2]$ \tablenotemark[2]         \\
\end{tabular}
\tablenotemark[1]{rank 0: gaussian with LLN, 1: gaussian with enhanced MSD, 2: distorted gaussian 
with enhanced MSD, 3: visible peaks on top of a broad band, 4: sharp peaks.}\\
\tablenotemark[2]{The correlator exponentially decays exhibiting the periodicity in the parenthesis. 
The dynamics can be described by unstable quasi-clusters in periodic motion.}\\
\tablenotemark[3]{The correlator does not decay and oscillates in period three. The periodic window.  }\\
\tablenotemark[4]{The correlator decays exponentially with longer life time than that at $\varepsilon=0$
but no low periodicity is observed. }\\
\tablenotemark[5]{A slanting surface between the first and the second-class peaks.}
\end{table}
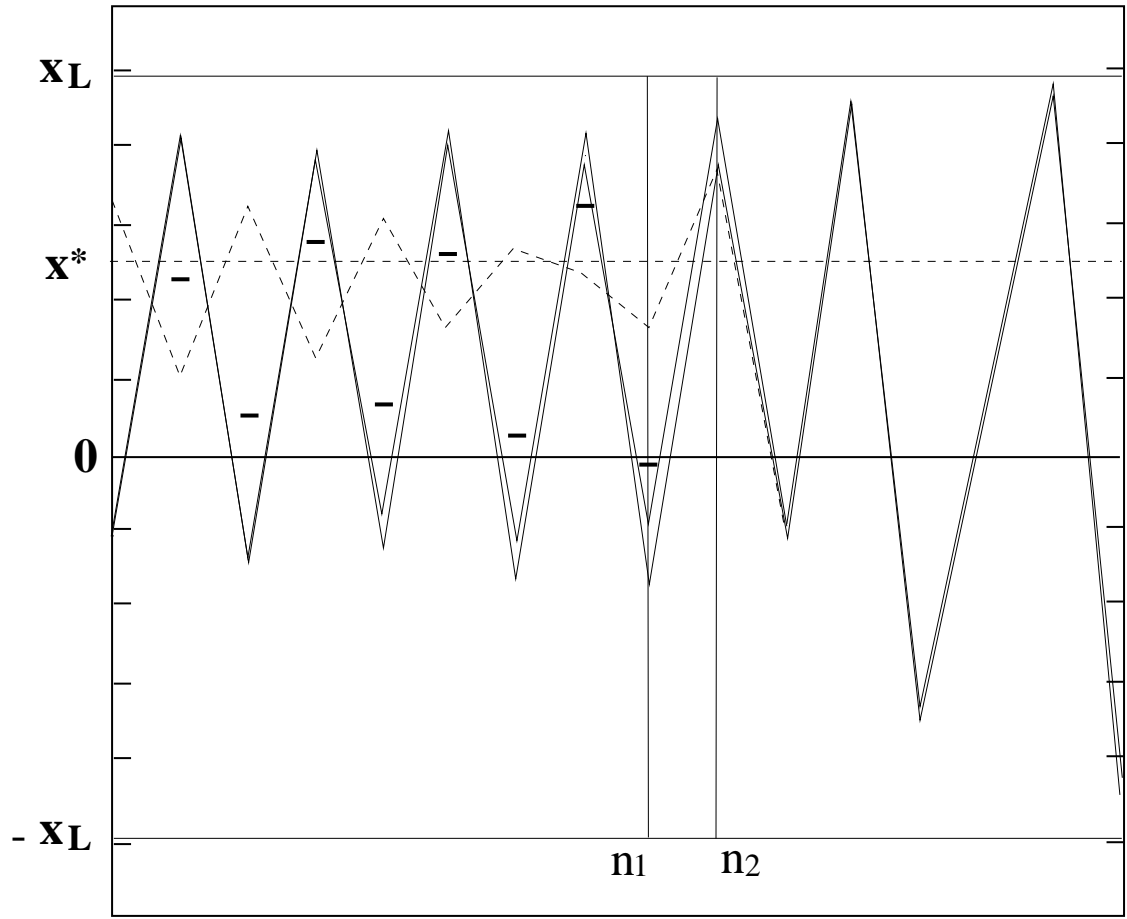
\end{document}